\documentclass[twocolumn,natbib,iop,hyperref]{aastex62}
\usepackage{natbib}

\bibpunct{(}{)}{;}{a}{}{,}
\usepackage{multirow}
\usepackage{color}
\usepackage[varg]{txfonts}

\newcommand{\msol}{M$_{\odot}$}

\newcommand{\cgs}{erg s$^{-1}$ cm$^{-2}$}
\newcommand{\pf}{\texttt{Platefit} }

\newcommand{\nii}{[\ion{N}{2}]}
\newcommand{\neiii}{[\ion{Ne}{3}]}
\newcommand{\hb}{H$\beta$}
\newcommand{\oiii}{[\ion{O}{3}]}
\newcommand{\oii}{[\ion{O}{2}]}
\newcommand{\ewoii}{EW$_{\mathrm{[O II]}}$}
\newcommand{\ha}{H$\alpha$}

\newcommand{\hd}{H$\delta$}
\newcommand{\hda}{H$\delta_{A}$}

\accepted{September 29, 2021}

\shorttitle{Ubiquitous \oii\ Emission in Quiescent Galaxies at $z\approx0.85$}
\shortauthors{Maseda et al.}

\begin{document}

\title{Ubiquitous \oii\ Emission in Quiescent Galaxies at $z\approx0.85$ from the LEGA-C Survey\footnote{Based on data products from observations made with ESO Telescopes at the La Silla Paranal Observatory under program ID 194.AF2005(A-N).}}
\correspondingauthor{Michael V. Maseda}
\email{maseda@astro.wisc.edu}

\author[0000-0003-0695-4414]{Michael V. Maseda}
\affil{Department of Astronomy, University of Wisconsin-Madison, 475 N. Charter Street, Madison, WI 53706, USA}
\affil{Leiden Observatory, Leiden University, P.O. Box 9513, 2300 RA, Leiden, The Netherlands}
\author[0000-0002-5027-0135]{Arjen van der Wel}
\affil{Sterrenkundig Observatorium, Universiteit Gent, Krijgslaan 281 S9, B-9000 Gent, Belgium}
\affil{Max-Planck-Institut f\"ur Astronomie, K\"onigstuhl 17, 69117 Heidelberg, Germany}
\author[0000-0002-8871-3026]{Marijn Franx}
\affil{Leiden Observatory, Leiden University, P.O. Box 9513, 2300 RA, Leiden, The Netherlands}
\author[0000-0002-5564-9873]{Eric F. Bell}
\affil{Department of Astronomy, University of Michigan, 1085 South University Ave., Ann Arbor, MI 48109, USA}
\author[0000-0001-5063-8254]{Rachel Bezanson}
\affil{Department of Physics and Astronomy, University of Pittsburgh, Pittsburgh, PA 15260, USA}
\author[0000-0002-9330-9108]{Adam Muzzin}
\affil{Department of Physics and Astronomy, York University, 4700 Keele St., Toronto, Ontario, Canada, MJ3 1P3}
\author[0000-0001-8823-4845]{David Sobral}
\affil{Department of Physics, Lancaster University, Lancaster LA1 4YB, UK}
\author[0000-0003-2388-8172]{Francesco D'Eugenio}
\affil{Sterrenkundig Observatorium, Universiteit Gent, Krijgslaan 281 S9, B-9000 Gent, Belgium}
\author[0000-0002-9656-1800]{Anna Gallazzi}
\affil{INAF-Osservatorio Astrofisico di Arcetri, Largo Enrico Fermi 5, I-50125 Firenze, Italy}
\author[0000-0002-2380-9801]{Anna de Graaff}
\affil{Leiden Observatory, Leiden University, P.O. Box 9513, 2300 RA, Leiden, The Netherlands}
\author[0000-0001-6755-1315]{Joel Leja}
\affil{Department of Astronomy \& Astrophysics, The Pennsylvania State University, University Park, PA 16802, USA}
\affil{Institute for Computational \& Data Sciences, The Pennsylvania State University, University Park, PA 16802, USA}
\affil{Institute for Gravitation and the Cosmos, The Pennsylvania State University, University Park, PA 16802, USA}
\author[0000-0001-5937-4590]{Caroline Straatman}
\affil{Sterrenkundig Observatorium, Universiteit Gent, Krijgslaan 281 S9, B-9000 Gent, Belgium}
\author[0000-0001-7160-3632]{Katherine E. Whitaker}
\affil{Department of Astronomy, University of Massachusetts, Amherst, MA 01003, USA}
\affil{Cosmic Dawn Center (DAWN), Niels Bohr Institute, University of Copenhagen, Juliane Maries vej 30, DK-2100 Copenhagen, Denmark}
\author[0000-0003-2919-7495]{Christina C. Williams}
\affil{Steward Observatory, University of Arizona, 933 North Cherry Avenue, Tucson, AZ 85721, USA}
\author[0000-0002-9665-0440]{Po-Feng Wu}
\affil{National Astronomical Observatory of Japan, Osawa 2-21-1, Mitaka, Tokyo 181-8588, Japan}
\affil{East Asia Core Observatory Association Fellow}

\begin{abstract}
Using deep rest-frame optical spectroscopy from the Large Early Galaxy Astrophysical Census (LEGA-C) survey, conducted using VIMOS on the ESO Very Large Telescope, we systematically search for low-ionization \oii\ $\lambda\lambda$3726,3729 emission in the spectra of a mass-complete sample of $z\approx0.85$ galaxies.  Intriguingly, we find that 59 percent of UVJ-quiescent (i.e. non star-forming) galaxies in the sample have ionized gas, as traced by \oii\ emission, detected above our completeness limit of 1.5 \AA.  The median stacked spectrum of the lowest equivalent width quiescent galaxies also shows \oii\ emission.  The overall fraction of sources with \oii\ above our equivalent width limit is comparable to what we find in the low-redshift Universe from GAMA and MASSIVE, except perhaps at the highest stellar masses ($>$ 10$^{11.5}$ \msol).  However, stacked spectra for the individual low-equivalent width systems uniquely indicates ubiquitous \oii\ emission in the higher-$z$ LEGA-C sample, with typical \oii\ luminosities per unit stellar mass that are a factor of $\times$3 larger than the lower-$z$ GAMA sample. Star formation in these otherwise quiescent galaxies could play a role in producing the \oii\ emission at higher-$z$, although it is unlikely to provide the bulk of the ionizing photons.  More work is required to fully quantify the contributions of evolved stellar populations or active galactic nuclei to the observed spectra.
\end{abstract}
\keywords{galaxies: elliptical and lenticular, cD --- galaxies: evolution}


\section{INTRODUCTION}

The traditional picture of ``quiescent" galaxies (i.e. galaxies with insignificant star formation given their stellar mass) is one of old stellar populations with little to no gas or dust in the interstellar medium.  These galaxies are often collectively referred to using morphological classifications such as ``early-type" or ``elliptical" as well as color-based classifications such as ``red."  Although they have formed the majority of their stars at early cosmic times \cite[$z>1$;][]{2005ApJ...633..174T}, many observations show that  they might not be truly ``red and dead," even many Gyr after the peak of their star formation activity.  Namely, many of these galaxies still contain reservoirs of ionized or neutral gas \cite[e.g.][]{1975ApJ...202....7G,1980AA....87..152H,2006MNRAS.366.1151S,2010AA...519A..40A,2012MNRAS.422.1835S,2016MNRAS.461.3111B,2018ApJ...860..103S}, although the exact frequency in the full population is not well known.  The presence of ionized gas in particular implies the presence of a source of ionizing photons in addition to a cool, old stellar population \citep{1969PASP...81..475C}.  This could be due to young stars from low levels of ongoing star formation and/or hot, evolved stars, with limited observational evidence supporting both possibilities.

In the case of young stars, recent episodes of star formation in otherwise quiescent galaxies are inferred based on e.g. mass-to-light ratios \citep{2002ApJ...564L..13T}, and in some cases ``rejuvenation" events are discovered, where the galaxy temporarily moves from the red sequence to the blue cloud \cite[e.g. in 16\% of quiescent galaxies at $z\approx0.8$;][see also \citeauthor{2021ApJ...907L...8A} \citeyear{2021ApJ...907L...8A}]{2019ApJ...877...48C}.  \citet{2007ApJS..173..512S} show that some star formation is occurring in 30\% of bright ($M_r < -21.5$) $0.05 < z < 0.10$ early-type galaxies, and \citet{2005ApJ...633..174T} show that the fractional contribution of secondary episodes of star formation increases with decreasing galaxy mass \cite[from $<$1\% for masses above 10$^{11.5}$ \msol\ to 20$-$40\% below 10$^{11}$ \msol; see also][]{2007ApJS..173..619K,2010MNRAS.404.1775T,2020NatAs...4..252S}.  

Theoretical models predict that evolved stars with ages greater than 100 Myr can also produce these ionizing photons \citep{1990ApJ...364...35G,1994AA...292...13B}.  Specifically, these are the broad classes of extreme horizontal branch (EHB) or post-asymptotic giant branch (post-AGB) stars, otherwise referred to as ``hot low-mass evolved stars'' \citep{2011MNRAS.413.1687C}.  Various models of composite stellar populations include different prescriptions for the duration and incidence rate of stars in these evolutionary phases, with wildly different predictions for the contribution to the total ionizing photon output \cite[e.g.][]{2006ApJ...652...85M,2010ApJ...722L..64K}. 

Many authors have used observations of the ``UV upturn" or ``UV-excess" as a way of identifying systems with copious amounts of ionizing photons \cite[][and references therein]{2008ASPC..392....3Y}.  This is an observed ``excess" of ultraviolet (UV) photons (1000$-$2500 \AA\ in the restframe) compared to the optical continuum produced by predominantly old stars.  Since the average stellar population is younger at higher-$z$ \citep{2014ApJ...788...72G}, galaxies will have fewer evolved low-mass stars.  Therefore, if these stars are the primary contributors to the UV luminosity then one would expect to see a decrease in the fraction of galaxies with a UV-excess with increasing redshift.   \citet{2016MNRAS.461..766L} and \citet{2018MNRAS.480.2236A} find this decrease out to $z \approx$ 0.7 in galaxies with stellar masses in excess of 10$^{11.5}$ \msol.  However, other authors have suggested that the UV photons do not need to come from ``exotic" stellar evolutionary phases (e.g. EHB or binaries), but instead can be produced by combinations of young and old stellar populations \citep{2020MNRAS.497.3251W}.  Indeed, \citet{2021MNRAS.500.1870D} find that ``passive" galaxies with a UV-excess have less recent star formation, are redder, and are more metal-rich. 

Problematically, using the UV alone is not enough to truly disentangle the contributions of recent star formation and evolved stellar populations \cite[e.g.][]{2000ApJ...541..126M}, particularly as the standard selection based on $NUV-r$ colors only isolate the most extreme cases of the UV upturn.  Spectroscopy is an important complement to photometric studies, where (low-ionization) emission lines such as \oii, \hb, or \ha\ also serve as signposts of ionized gas.  \citet{2018MNRAS.481.1774H} differentiate between early-type galaxies with and without emission lines at $z < 0.095$, finding that the excess UV luminosity in the former can be attributed to more recent star formation (0.1$-$5 Gyr) even though both classes have similar levels of ionizing photon production from their old stars \cite[see also][]{2011MNRAS.413.1687C}.  Taken together with the decreased incidence of UV-upturn galaxies, this would imply that more recent star formation is occurring in the quiescent population at higher-$z$.  This in turn implies a predicted increased incidence rate of ``liny" quiescent galaxies at higher-$z$.

Despite the crucial additional information that spectroscopy can provide, it is more difficult to obtain due to the required S/N values.  Moreover, emission lines produced by extremely low levels of star formation or by evolved stars are predicted to have low equivalent widths (EWs), typically $\lesssim$ 5 $\mathrm{\AA}$ depending on the line \cite[cf. the \oii\ detection threshold of $\approx$ 5 $\mathrm{\AA}$ in][see also the discussion in \citeauthor{2013AA...558A..61M} \citeyear{2013AA...558A..61M}]{1999ApJS..122...51D,2002ApJ...564L..13T,2010ApJ...716..970L,2017ApJ...850..181R,2017ApJ...838...94W}.  Hence, many current spectroscopic studies of quiescent galaxies are typically restricted to the low-$z$ Universe \cite[e.g. $z\lesssim0.1$][]{2018MNRAS.481.1774H,2020MNRAS.497.3251W}.   At higher redshifts they are mostly limited to small samples \cite[e.g. $<$ 10 objects;][]{2003ApJ...584L..69B,2013ApJ...764L...8B,2014ApJ...788...72G,2018ApJ...862..125N,2018AA...618A..85S}, rely on spectral stacking to obtain an averaged result for all the galaxies in the sample \citep{2017AA...599A..95G}, and/or only probe the most massive galaxies \cite[e.g. M$_{\star}~>$ 10$^{11.5}$ \msol;][]{2016MNRAS.461..766L}.  Selection effects also hamper simple interpretations of the results since these samples are also selected heterogeneously: \citet{2018MNRAS.481.1774H} use a morphological criteria in combination with a cut at $EW_{\mathrm{H\alpha}}~<$ 3 \AA; \citet{2020MNRAS.497.3251W} use the same morphological cut but also cut in $NUV-r$ color; \citet{2017ApJ...838...94W} utilize a different morphological cut in addition to a cut in specific star formation rate; \citet{2010ApJ...716..970L} cut in $i'-z'$ versus $z'$; and \citet{2016MNRAS.461..766L} cut in $g-i$ color and UV-derived stellar age.

The first step in understanding the ionizing source(s) in these galaxies is understanding the prevalence of emission lines in a homogeneously-selected sample.  Once the incidence rate is established, analyzing the variation in those signatures with other properties (such as the total stellar mass of the galaxy or the spatial distribution of the features) can help to determine the relative contributions of different production mechanisms.  Here, we establish the incidence rate of the low-ionization \oii\ emission line out to $z\approx1$.  \oii\ in particular is the key emission line to study since it is observed to trace more massive galaxies and recovers more of the ``less active" population of galaxies at $z \gtrsim$ 1 compared to Balmer lines \citep{2016MNRAS.463.2363K}.  Deep, high-resolution spectroscopy is required to accurately measure \oii\ (and \hb) at low EWs, where Balmer absorption features can significantly impact the observed line flux and hence a high continuum S/N is necessary.

The advent of deep spectroscopic surveys, like the Large Early Galaxy Astrophysical Census (LEGA-C) survey conducted with the VIMOS instrument on the ESO Very Large Telescope \citep{2016ApJS..223...29V}, now allows us to measure faint emission lines systematically in representative populations of galaxies at moderate redshifts.  In particular, LEGA-C offers high continuum S/N ($\approx$ 20 \AA$^{-1}$), high spectral resolution ($R \approx$ 3500), and spectral coverage of \oii\ over the redshift range 0.6 $< z <$ 1.1 for a mass-complete sample of galaxies.  This redshift range, when the Universe was approximately half of its current age, is  when the population of massive quiescent galaxies observed locally is building up most rapidly \citep{2004ApJ...608..752B,2007ApJ...665..265F}.  LEGA-C therefore presents us with a unique opportunity to assess the strength and prevalence of \oii\ emission with the aim of understanding the prevalence of ionizing sources in these apparently ``quiescent" systems.

This paper is organized as follows.  In Section \ref{sec:data} we introduce our sample, drawn from the LEGA-C survey, and describe how we separate star-forming from quiescent galaxies.  We describe our methodology for fitting emission lines like \oii.  We present the resulting sample and discuss the relevance and caveats in Section \ref{sec:sample} and compare our results to those derived from the GAMA survey of the low-$z$ Universe in Section \ref{sec:gama}.  Finally, we make concluding remarks and outlining future work in Section \ref{sec:conclusions}.  We adopt a flat $\Lambda$CDM cosmology ($\Omega_m=0.3$, $\Omega_\Lambda=0.7$, and H$_0=70 ~$km s$^{-1}$ Mpc$^{-1}$), AB magnitudes \citep{1974ApJS...27...21O}, and a \citet{2003PASP..115..763C} initial mass function throughout.  We also adopt the convention that emission line EWs (measured in $\mathrm{\AA}$) are positive.

\section{Data}
\label{sec:data}

Our spectra come from the third data release (DR3) of the LEGA-C survey (A. van der Wel et al., ApJS in press.), containing 4081 spectra of 3741 individual galaxies.  We remove any galaxies with significant X-ray or radio fluxes \citep{2016ApJ...819...62C,2017ApJ...847...72B} as they could plausibly have spectral contamination from an active galactic nucleus (AGN).  As the spectral coverage varies across the VIMOS field-of-view, requiring spectral coverage of \oii\ does not directly translate into a redshift selection.  The 2019 unique galaxies in the $K_s$-selected (mass-complete) sample with spectral coverage of \oii\ range from $z$ = 0.57 $-$ 1.44, with a median $\left<z\right>$ =  0.85.

To separate our sample into star-forming and quiescent galaxies, we utilize the UVJ selection of \citet{2011ApJ...735...86W}.  This selection differentiates between old stellar populations and dust extinction by using two (restframe) colors: ($U~-~V$) and ($V~-~J$).  We determine the restframe colors for the LEGA-C galaxies using their spectroscopic redshifts and the results from the \texttt{EAZY} code \citep{2008ApJ...686.1503B} applied to the UltraVISTA DR4 photometric catalog \citep{2012AA...544A.156M,2013ApJS..206....8M}.  Applying the UVJ selection yields 843 quiescent galaxies and 1176 star-forming galaxies.  Using \texttt{Prospector} \citep{2017ApJ...837..170L,2019ascl.soft05025J} to fit the observed $B$ through $J$ plus 24 $\mathrm{\mu}$m photometry from \citet{2013ApJS..206....8M} with revised zero points (A. van der Wel et al. submitted), we find that the median specific star formation rates (star formation rate per unit stellar mass; sSFR) are 10$^{-11.9}$ yr$^{-1}$ for the quiescent galaxies and 10$^{-9.6}$ yr$^{-1}$ for the star-forming galaxies.

\subsection{Measurements of \oii}
\label{sec:pf}

\begin{figure*}
\begin{center}
\includegraphics[width=.95\textwidth]{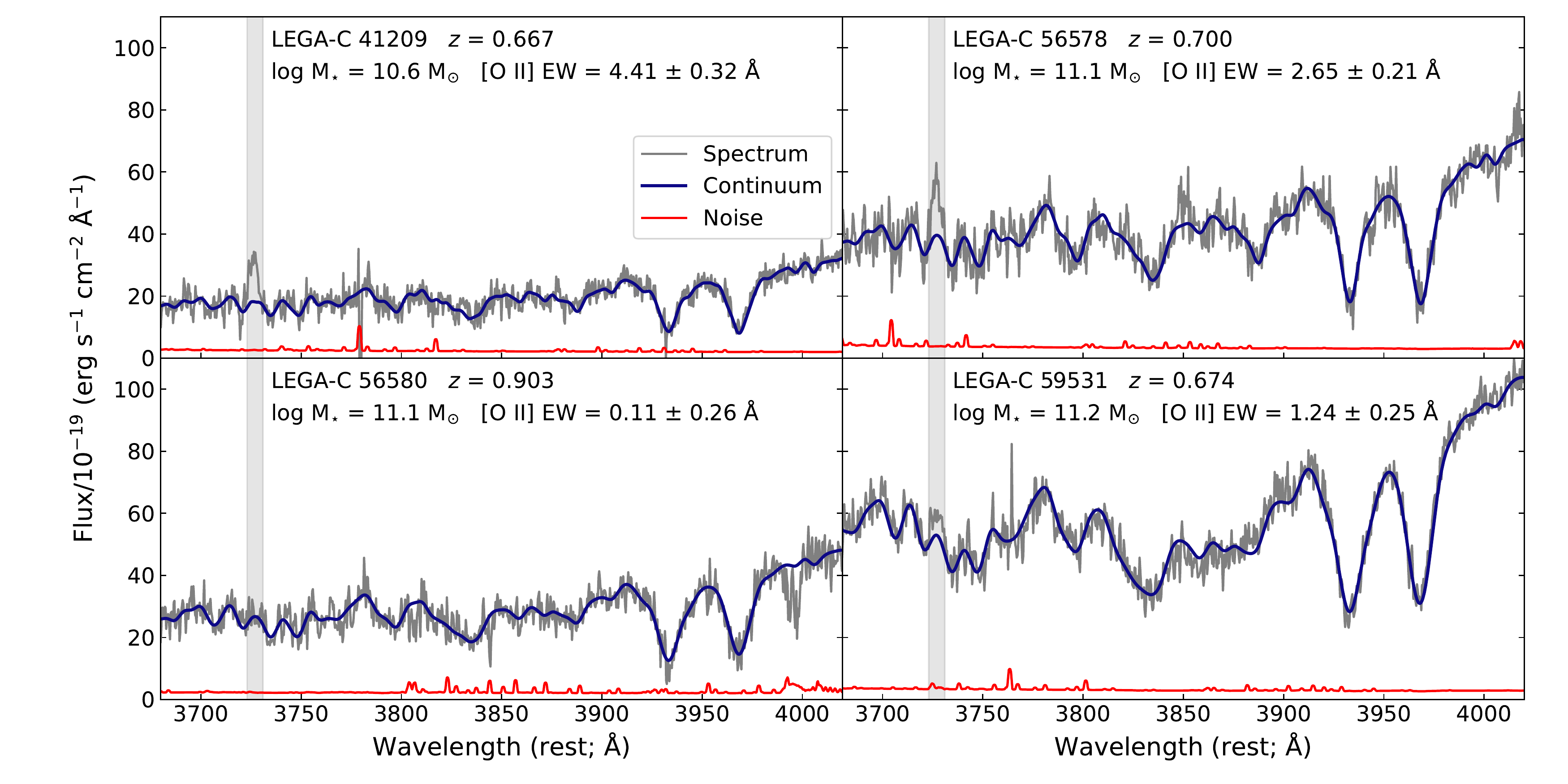} 
\caption{Representative LEGA-C spectra of UVJ-quiescent galaxies, where the gray line shows the 1D spectrum, the red line shows the 1-$\sigma$ uncertainties, the blue line shows the best-fit \texttt{pPXF} stellar continuum, and the shaded gray region highlights the expected position of \oii.  The top two panels are classified as detections of \oii\ due to their EW values that are in excess of 1.5 \AA.  The bottom two panels are ``non-detections'' since their EWs are below this threshold; although the bottom-right panel still has \oii\ detected at nearly 5-$\sigma$, this is below our completeness limit and hence we would not have been able to detect such a low-EW line in all galaxies in the survey (see Section \ref{sec:sample}).  The typical velocity width of the emission lines ($\approx$ 200 km s$^{-1}$) means that the \oii\ doublet in the quiescent sample is typically unresolved at spectral resolution of VIMOS.}
\label{fig:specs}
\end{center}
\end{figure*}

We must model the stellar continuum and its associated absorption features while fitting \oii\ since the line is close in wavelength to high-order Balmer absorption lines (i.e. H13 at 3734.37 \AA).  To do so, we fit the stellar continuum for each galaxy using the \texttt{pPXF} code \citep{2004PASP..116..138C,2012ascl.soft10002C,2017MNRAS.466..798C}, fixed to the LEGA-C redshift.  The stellar continuum is modeled with the empirical stellar population templates of \citet{2010MNRAS.404.1639V}\footnote{Our results are qualitatively similar if we instead choose to use other stellar population models: see Section \ref{sec:sps}.}.  Further details of the stellar continuum fits can be found in \citet{2016ApJS..223...29V}, \citet{2018ApJ...868L..36B}, and \citet{2018ApJS..239...27S}.  Example spectra, including the best-fit stellar continuum model, are shown in Figure \ref{fig:specs}.  After fitting the stellar continuum, we subtract the best-fit model from the 1D spectrum and run \pf \citep{2004ApJ...613..898T, 2004MNRAS.351.1151B,2020MNRAS.tmp.2359G} to simultaneously fit the emission lines, again fixed to the LEGA-C redshift but allowing for a $\pm$ 300 km s$^{-1}$ velocity offset between the Balmer lines and the forbidden lines (such as \oii).  Equivalent widths are calculated based on the measured line flux and the integrated continuum flux from the continuum model over the same wavelength range.  Throughout this work all EWs are given in the rest-frame.

We determine the uncertainties on the fits by perturbing the flux at each wavelength position in the spectrum according to the noise level at that position and re-fitting.  We repeat this procedure 100 times per spectrum and derive the line flux uncertainties according to the standard deviation of the individual fits.  While this procedure determines the formal fitting uncertainty to the line flux (and EW), uncertainties in the overall flux calibration and in the continuum subtraction are not included.  In order to establish the magnitude of these contributions to our ability to detect \oii\ emission, we analyze 27 objects with duplicate spectra covering \oii\ using the procedure of \citet{2008AA...485..657B}.  These objects were selected to be quiescent according to UVJ, with measured \oii\ EWs $<$ 1.5 \AA\ (see Section \ref{sec:sample}).  The uncertainty in both the \oii\ flux and EW is a factor of 1.47 higher than is indicated by the formal fitting uncertainties.  Therefore, we increase the error in both line flux and EW by this amount.  We note that this added uncertainty could be due to repeat observations covering slightly different physical regions of the galaxy, although this effect is typically less than 0.2 arcseconds (1.5 kpc at $z=0.8$; J. van Houdt et al. submitted).  For objects with duplicate spectra, we take the weighted mean of the two independent measurements, including this factor of 1.47 to account for systematics.

\section{The LEGA-C Sample of \oii\ Emitters}
\label{sec:sample}
In order to define a ``detection" of \oii, we must first determine our completeness as a function of line EW.  The use of an EW instead of line luminosity to define completeness is motivated by the difference in rest-frame $U$-band continuum luminosities probed within the observed-frame $K_s$-band selected LEGA-C sample.  At a fixed line luminosity we are sensitive to much lower EWs in sources that are brighter in the continuum.  Therefore, we define our completeness limit for the sample of \oii\ emitters to be 1.5 \AA, restframe.  An EW of 1.5 \AA\ corresponds to the theoretical expectation for a 3-$\sigma$ detection of the combined \oii\ doublet based on our line fitting technique described above at the faintest $U$-band continuum magnitude of the primary LEGA-C sample ($m_U\approx$ 25).  In practice, we find that 96 (90) percent of LEGA-C (quiescent) galaxies with \oii\ EWs above this threshold are indeed measured to have  a significance greater than 3-$\sigma$.  For quiescent galaxies with EWs below this threshold, only 16 percent have a significance greater than 3-$\sigma$, justifying this demarcation.

Applying this cut leaves us with a sample of 1644 galaxies with \oii\ EWs above 1.5 \AA: 1151 star-forming and 493 quiescent corresponding to detection fractions of 0.979 and 0.585, respectively.  When applying a correction factor to account for the spectroscopic completeness of the LEGA-C survey compared to the parent UltraVISTA catalog (``Scor;'' van der Wel et al. in press), the overall detection fractions are 0.982 for star-forming galaxies and 0.594 for quiescent galaxies: the majority of galaxies in the LEGA-C survey, both star-forming and quiescent, have detections of \oii\ emission above 1.5 \AA\ equivalent width.  

\begin{figure}
\begin{center}
\includegraphics[width=.45\textwidth]{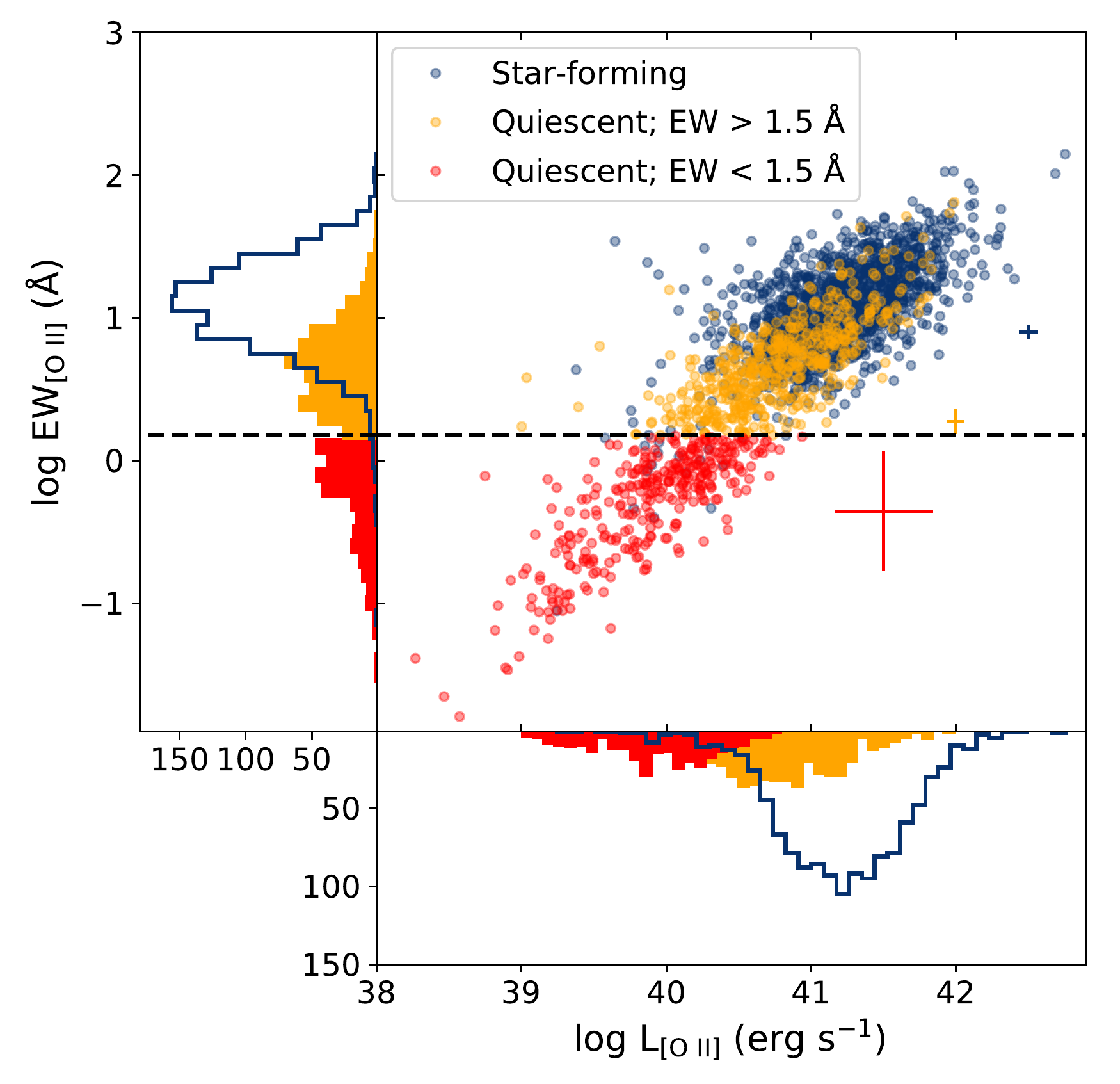} 
\caption{\oii\ luminosity versus (restframe) EW for LEGA-C quiescent and star-forming galaxies, based on the UVJ diagram.  The horizontal dashed line represents the EW threshold of 1.5 \AA\ used to differentiate between ``detections'' (orange) and ``non-detections'' (red).  Individual sources with EWs below our threshold are plotted at their bootstrap median value (see Section \ref{sec:pf}), with 46 galaxies (5 \%) not shown as they have negative \oii\ luminosities and/or EWs.  Characteristic uncertainties are shown for each subsample.}
\label{fig:hists}
\end{center}
\end{figure}

\begin{figure*}[htbp]
\begin{center}
\includegraphics[width=.63\textwidth]{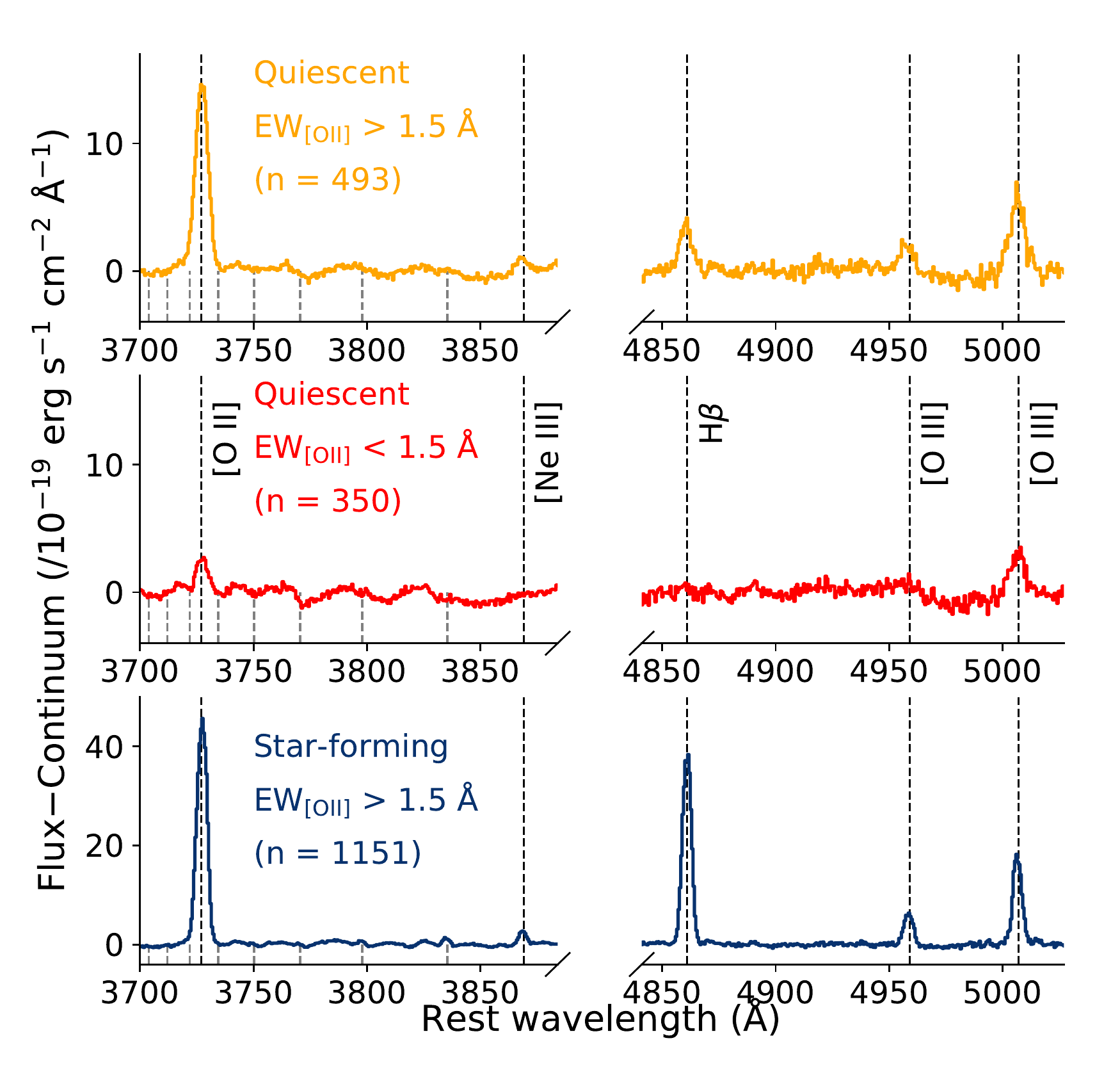}
\includegraphics[width=.27\textwidth]{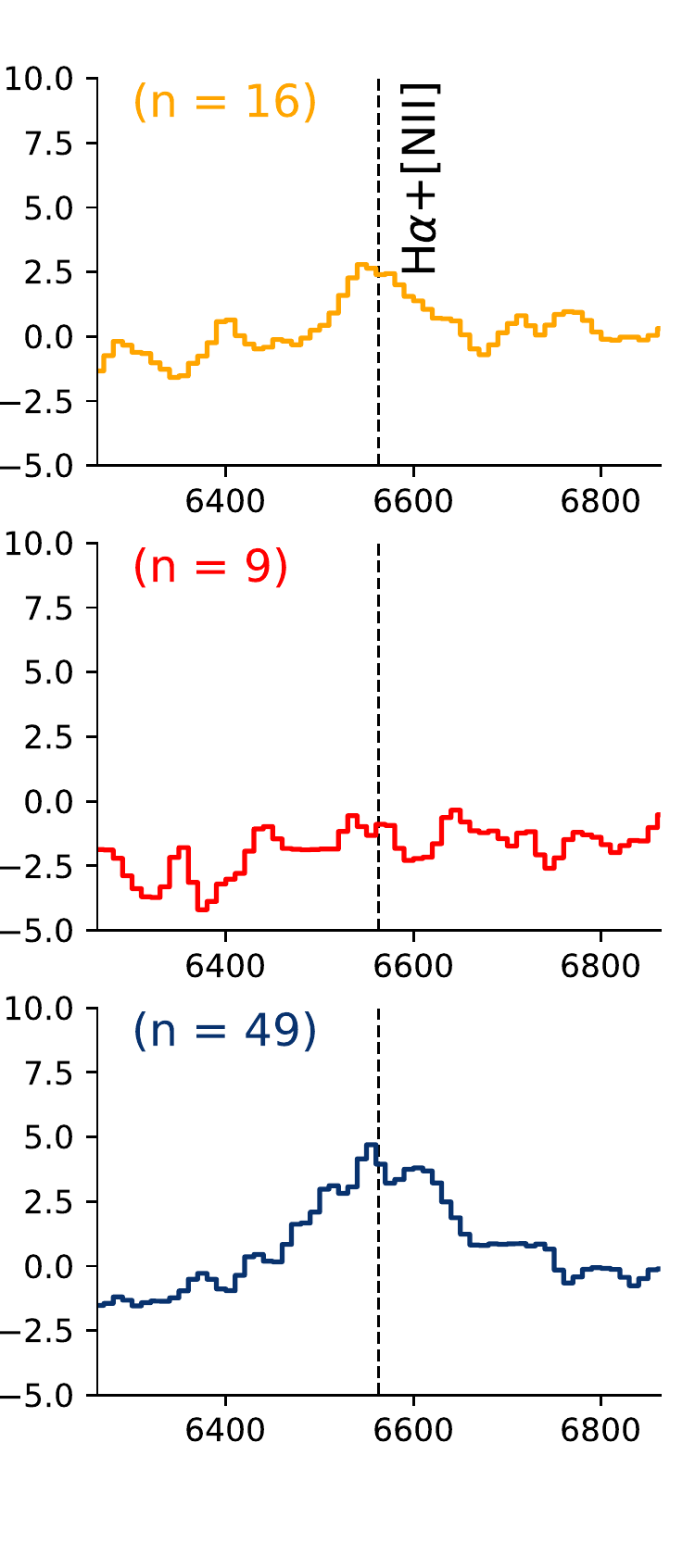}
\caption{Median continuum-subtracted stacks for star-forming galaxies (bottom) and quiescent galaxies with (top) and without (middle) \oii\ emission in excess of 1.5 \AA.  The positions of prominent emission lines are denoted by dashed lines.  Balmer absorption features are denoted by the short gray dashed lines.  The left and center panels show stacks using the LEGA-C VIMOS data, and the right panels show stacks using HST/WFC3 G141 data.  Both quiescent galaxy stacks show prominent \oii\ emission, even when it is not detected in individual spectra (middle-left panel) indicating that this feature is ubiquitous in the LEGA-C sample.}

\label{fig:stacks}
\end{center}
\end{figure*}

In Figure \ref{fig:hists} we show the measured \oii\ luminosities and EWs for the LEGA-C sample, split according to their classification from the UVJ diagram.  Star-forming galaxies have, on average, more luminous \oii\ emission and a higher \oii\ EW than their quiescent counterparts: the median \oii\ EW for the quiescent galaxies is 2.4 \AA\ compared to 11.8 \AA\ for the star-forming galaxies.  Figure \ref{fig:stacks} shows stacked spectra for UVJ-quiescent galaxies with (top panel) and without (middle panel) \oii\ emission in excess of 1.5 \AA, and for star-forming galaxies (bottom panel).  The individual spectra are all continuum-subtracted (as described in Section \ref{sec:pf}) before stacking.  In both cases, the median spectra show clear detections of \oii\ and \oiii.  In the case of objects without individual detections of \oii, the stack has a measured EW of 0.59 \AA; see Table \ref{tab:lines}.  This implies that the average UVJ-quiescent galaxy has low-levels of \oii\ (and \oiii) emission, even though it can only be detected individually at EWs $>$ 1.5 \AA\ with the typical S/N of the LEGA-C VIMOS data (see also Section \ref{sec:ew} and \citeauthor{2017ApJ...838...94W} \citeyear{2017ApJ...838...94W}).  

The right panels of Figure \ref{fig:stacks} show stacked HST/WFC3 G141 spectra from the 3D-HST Survey \citep{2012ApJS..200...13B,2016ApJS..225...27M} for the subset of LEGA-C targets in the same footprint.  The best-fit stellar continuum model as described in \citet{2016ApJS..225...27M} is subtracted from the extracted 1D spectrum, as is the case for the VIMOS spectra.  The stacked S/N is considerably lower than the LEGA-C VIMOS stacks due to lower individual S/N in the spectra and many fewer objects, but in the case of the star-forming galaxies and in the quiescent galaxies with individual detections of \oii\ there are clear detections of \ha $+$\nii\ \cite[unresolved at the G141 spectral resolution; see also][]{2013ApJ...770L..39W,2016ApJ...822....1F,2020MNRAS.491.2822C}.

\begin{deluxetable}{lcc}
\tablecaption{Measurements of emission line strengths (flux and EW) from the quiescent galaxy stacks shown in Figure \ref{fig:stacks}.  Error bars are determined using bootstrap resampling of the input spectra.\label{tab:lines}}
\tablehead{\colhead{Line} & \colhead{\ewoii\ $>$ 1.5 \AA} & \colhead{\ewoii\ $<$ 1.5 \AA} }
\startdata
\cutinhead{EW ($\mathrm{\AA}$)}
\oii & 4.6 $\pm$ 0.15 & 0.59 $\pm$ 0.052\\
\neiii & 0.16 $\pm$ 0.062 & $<$ 0.063\\
\hb & 0.39 $\pm$ 0.093 & $<$ 0.39 \\
\oiii & 0.51 $\pm$ 0.17 & 0.24 $\pm$ 0.037\\
\ha\ + \nii & 8.4 $\pm$ 1.1 & $<$ 2.8\\
\cutinhead{Flux (10$^{-19}$ \cgs)}
\oii & 100. $\pm$ 5.1 & 15. $\pm$ 2.3\\
\neiii & 4.4 $\pm$ 1.1 & $<$ 5.8\\
\hb & 20. $\pm$ 2.8 & $<$ 21.\\
\oiii & 35. $\pm$ 5.0 & 15. $\pm$ 5.9\\
\ha\ + \nii & 270 $\pm$ 27. & $<$ 150\\
\enddata
\tablecomments{When given, upper limits are 3-$\sigma$.}
\end{deluxetable}

\subsection{The selection of quiescent galaxies}
\label{sec:uvj}

As described in Section \ref{sec:data}, we utilize the UVJ diagram to differentiate between star formation and quiescence in the LEGA-C population.  This selection was developed in order to differentiate red spectral energy distributions due to dust attenuation from those due to intrinsically old stellar populations \citep{2007ApJ...655...51W,2009ApJ...691.1879W}.  As one of the potential interpretations of the presence of \oii\ emission is ongoing star formation, we must question if the UVJ selection is accurately separating the star-forming from the most quiescent galaxies: any mis-identifications of star-forming galaxies as quiescent galaxies would lead to elevated detection fractions of \oii\ in the ``quiescent" sample.

\begin{figure*}
\begin{center}
\includegraphics[width=.95\textwidth]{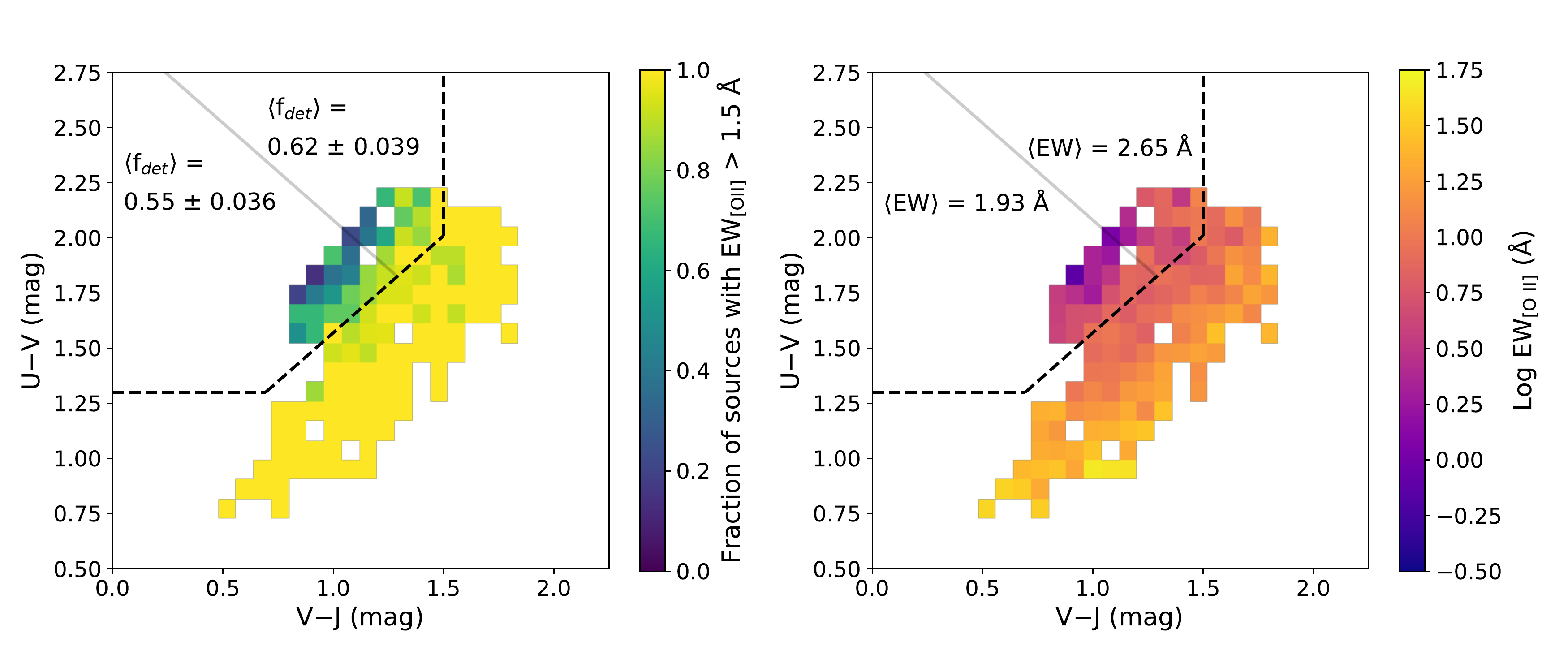}
\caption{Restframe $U-V$ and $V-J$ colors from \texttt{EAZY} \citep{2008ApJ...686.1503B} for the LEGA-C sample color-coded by the fraction of sources with \oii\ EW $>$ 1.5 \AA\ (left) and \oii\ EW (right) for all galaxies within each bin.  The dividing line between quiescent and star-forming from \citet{2011ApJ...735...86W} is denoted with the dashed lines.  The deep UltraVISTA photometry results in small formal errors in the calculated rest-frame colors (typically less than 0.1 magnitudes), and hence photometric scatter cannot be the primary cause of the high \oii\ detection fractions in color-selected quiescent galaxies.  When splitting the sample into to bins with equal numbers of sources (solid line), we observe that galaxies with redder $U-V$ and $V-J$ colors have systematically larger detection fractions and median EWs (text labels).}
\label{fig:uvj}
\end{center}
\end{figure*}

In Figure \ref{fig:uvj}, we show the UVJ diagram color-coded by the average \oii\ luminosity (left) and EW (right) for galaxies within a bin in restframe colors.  The  fraction of \oii\ emission above 1.5 \AA mimics the separation between quiescent and star-forming galaxies in UVJ-space, where the vast majority (94 percent) of cases with low-EW \oii\ lie in the quiescent region \cite[see also][]{2016MNRAS.463.2363K}.  The deep, multi-wavelength photometry also means that formal uncertainties in the rest-frame $U$, $V$, and $J$ photometry are also small: in the case of LEGA-C, random scatter from photometric errors can contribute only up to 0.5 percent to the overall detection fraction of \oii\ in quiescent galaxies.  For $z < 0.09$ from GAMA (see Section \ref{sec:gama}), this is 0.9 percent.  Only 1.8 percent of all star-forming galaxies according to the UVJ diagram do not have \oii\ detected above 1.5 $\mathrm{\AA}$, and even there we see the presence of \oii\ in the stacked spectrum at an EW of 1.5 $\mathrm{\AA}$, implying that sensitivity is the primary cause for the individual ``non-detections.''  We find that the UVJ star-forming criteria accurately identifies galaxies with \oii\ emission, plausibly from ongoing star-formation.  

When separating the quiescent galaxies into redder and bluer subsets in UVJ-space, we see that the redder sources have higher detection fractions (62\% compared to 55\%) and higher median \oii\ EWs (2.65 \AA\ compared to 1.93 \AA) than their bluer counterparts.  \citet{2013ApJ...770L..39W} use this same type of selection as a way of splitting quiescent galaxies into older (redder; $\approx$ 1.6 Gyr) and younger (bluer; $\approx$ 0.9 Gyr) populations at $1.4 < z < 2.2$ \cite[see also][]{2010ApJ...719.1715W}.  They find that the older/redder population has larger \oiii\ and \hb\ EWs, similar to what we observe in \oii.  However, the relationship between age and emission line properties from \citet{2013ApJ...770L..39W} is more binary, namely that they do not detect any \oiii\ or \hb\ emission in their stack of younger galaxies whereas we still observe \oii\ in our equivalently-selected systems.  This could be due to their low spectral resolution ($R\sim$130) or a genuine evolution in the ionizing properties of quiescent galaxies with redshift (see Section \ref{sec:discussion}), however without spectral coverage of \oii\ in the \citet{2013ApJ...770L..39W} sample it is difficult to draw a strong conclusion.

Figure \ref{fig:uvj} also shows indications that the EW of \oii\ emission varies systematically with perpendicular distance from the diagonal dividing line $UV = 0.88 \times VJ + 0.69$.  This matches the predicted trend of specific star formation rate (sSFR) from \citet{2019ApJ...880L...9L}, where the lowest-sSFR galaxies are furthest above this dividing line and vice-versa, although they also note that the UVJ colors ``saturate" for the sSFR values below 10$^{-10.5}$ yr$^{-1}$.  This is in excess of the median value we observe for the LEGA-C quiescent galaxies (10$^{-11.9}$ yr$^{-1}$).  Further work is required to assess if \oii\ EW, rather than UVJ colors, is a more efficient predictor of sSFR or total amount of ``quiescence" in the LEGA-C survey.

\subsection{Systematic differences between low and high-EWs}
\label{sec:diffs}

\begin{figure}
\begin{center}
\includegraphics[width=.45\textwidth]{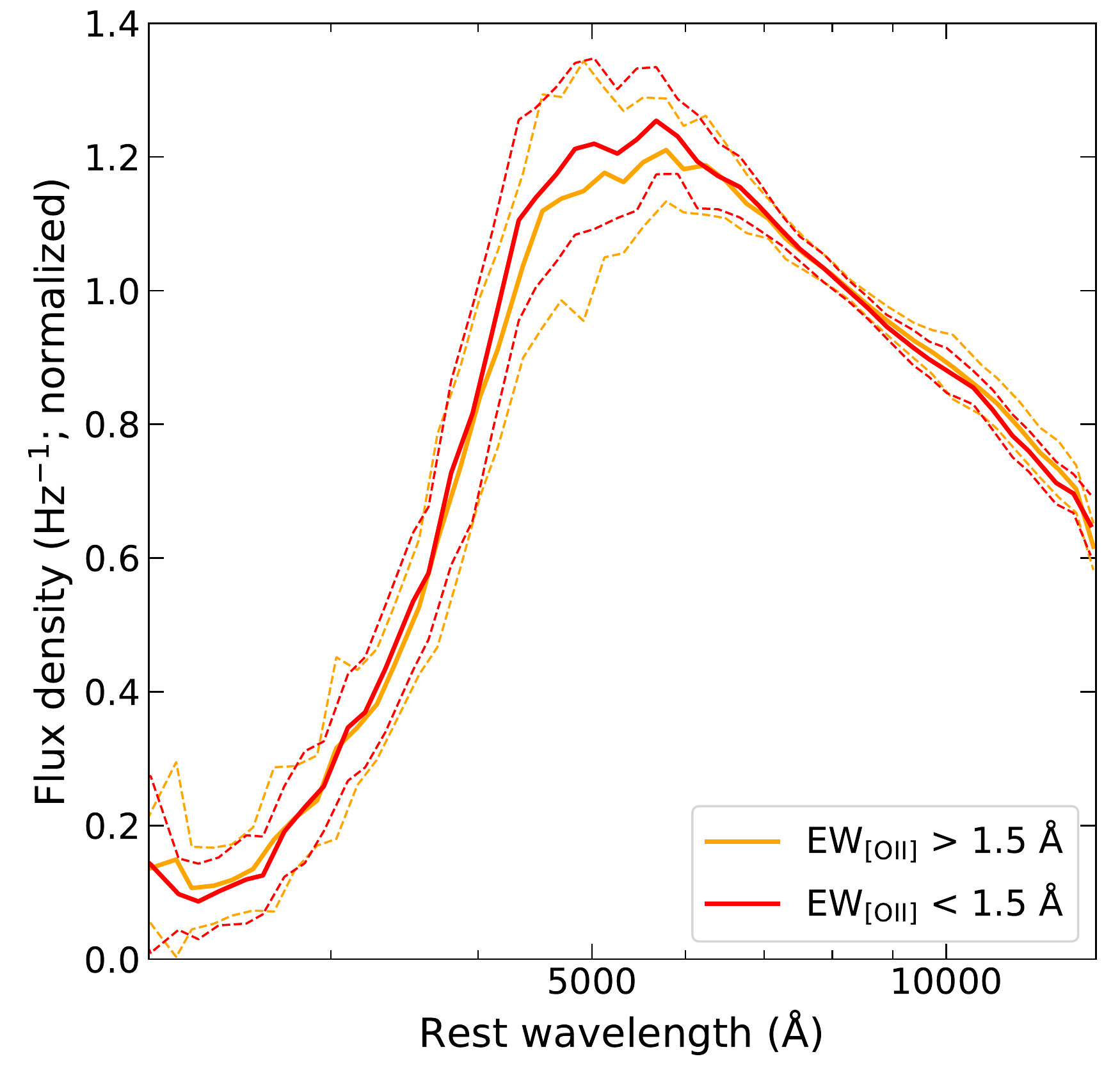} 
\caption{Composite SED for the two samples of quiescent galaxies (with and without \oii\ detections above our completeness limit), normalized at 8000 \AA.  Solid lines show the median SED using the observed optical/near-IR photometry, and the dashed lines show the extent of bootstrap resampling of the input sample \cite[see][]{2010ApJ...722L..64K}.  The SEDs of the two populations overlap significantly, implying that they have broadly similar stellar populations.  Purely based on the observed SEDs, we could not have consistently selected \oii\ emitters.}
\label{fig:sed}
\end{center}
\end{figure}

Beyond the UVJ colors, we can assess if the overall spectral energy distributions (SEDs) are similar between the high and low EW subsamples.  Figure \ref{fig:sed} shows the composite restframe SEDs for the two samples of quiescent galaxies, spanning the observed $B$-band to $K_s$-band.  The composite SEDs are created by normalizing the observed photometry to the (interpolated) restframe flux at 8000 \AA, and the resulting running average flux is recorded for the sample \cite[see][]{2010ApJ...722L..64K}.  A bootstrap resampling of the samples show the distribution in these composites (dashed lines).  Overall, the SEDs for the high and low EW samples are comparable in the wavelength regime probed here \cite[cf. GALEX/near-UV detections in some quiescent galaxies from][]{2007ApJS..173..512S}, with the average spread within the samples (dashed lines) being larger than the differences between the two samples.  \citet{2014ApJ...796...35F} found the same result at $0.3 < z < 2.5$  for UVJ-quiescent galaxies with and without 24 micron detections.  Thus, once an SED can be classified as ``quiescent'' there are no further signatures that are indicators of the relative strength of \oii\ emission.

\begin{figure}
\begin{center}
\includegraphics[width=.45\textwidth]{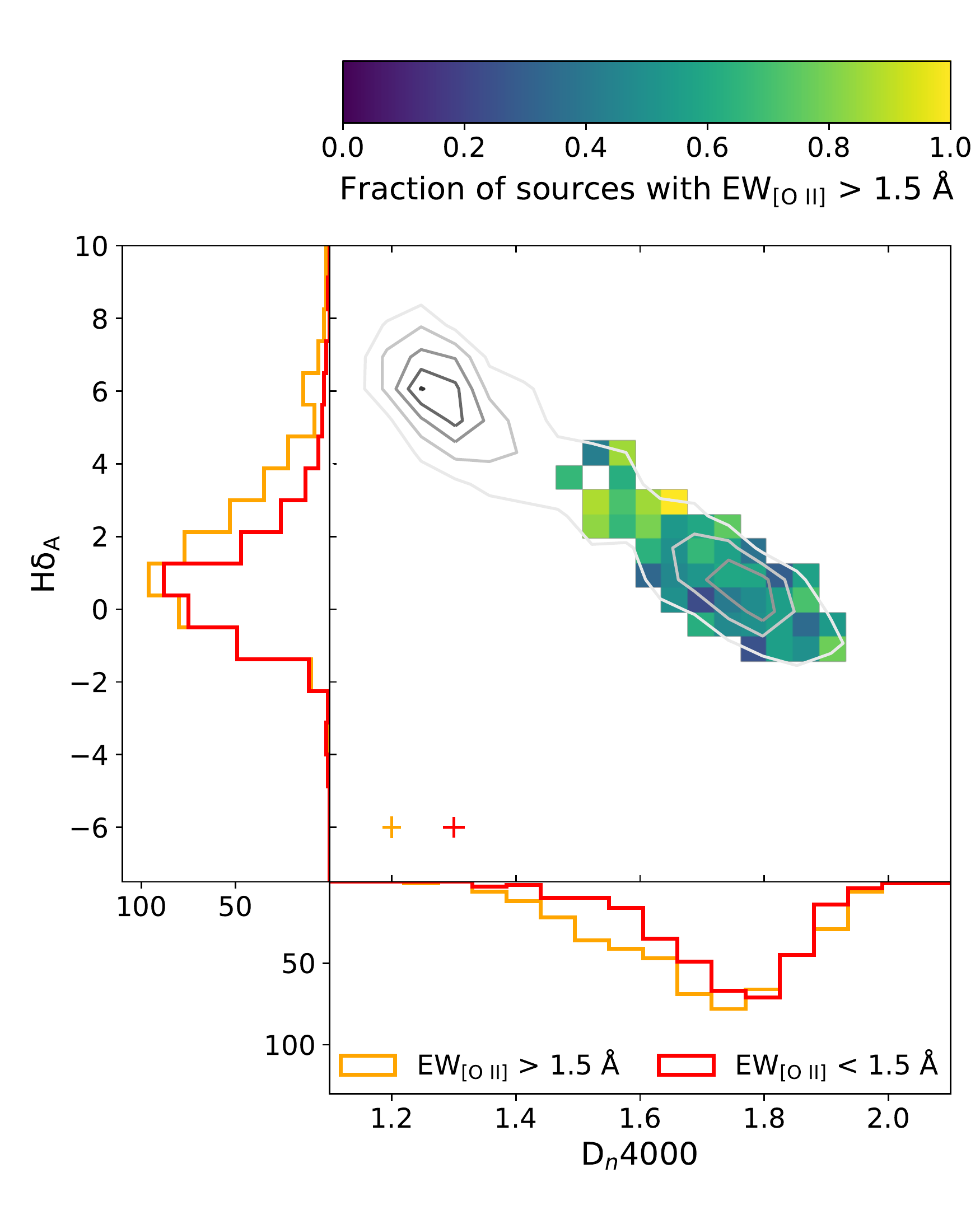} 
\caption{The detection fraction of \oii\ for LEGA-C quiescent galaxies as a function of D$_n$4000 and \hda, which together are sensitive to the age of the stellar population (for clarity, only bins with more than five galaxies are shown).  Contours (grayscale) denote the distribution of D$_n$4000 and \hda\ for all galaxies in LEGA-C (quiescent and star-forming).  The detection fraction increases as galaxies become younger, i.e. towards the upper-left of the diagram.  Furthermore, the individual distributions of D$_n$4000 and \hda\ are significantly different for the high- and low-EW subsamples (3.8-$\sigma$ and 4.8-$\sigma$, respectively).}
\label{fig:d4000}
\end{center}
\end{figure}

When looking at the D$_n$4000 and \hda\ spectral indices we see differences between the samples.  Figure \ref{fig:d4000} illustrates that the \oii\ detection fraction decreases in galaxies with larger D$_n$4000 and weaker \hd\ absorption, which corresponds to an older stellar population.  This difference compared to the behavior seen in splitting galaxies according to redder/bluer UVJ colors, where redder galaxies have a larger median EW and detection fraction (Figure \ref{fig:uvj}), can be attributed to dust attenuation: dust systematically reddens the spectra of the higher-EW galaxies at a fixed age, as measured by D$_n$4000 and \hd\ \cite[see also][]{2018MNRAS.481.1774H}.  Qualitatively, much like the separation on UVJ colors into redder and bluer sources, there are non-zero \oii\ detection fractions in all parts of the parameter space covered by quiescent galaxies.  The marginally larger difference between the \hda\ distributions compared to D$_n$4000 (4.8-$\sigma$ compared to 3.8-$\sigma$) does indicate that star formation could potentially be playing a role in the \oii\ production, as the \hda\ index evolves more strongly with age than D$_n$4000 on timescales $\lesssim$ 1 Gyr \citep{2003MNRAS.341...33K}.  We discuss this possibility further in Section \ref{sec:gama}, but we stress that accurate measurements of these spectral quantities is only possible with the high S/N afforded by the LEGA-C survey.

\subsection{The detection fraction of \oii\ in LEGA-C}
With the sample of quiescent galaxies established by UVJ, we now turn our attention to effects from our analysis procedure on the derived fraction of \oii\ emitters.

\subsubsection{Impact of the EW selection}
\label{sec:ew}

Based on the spectral stacking shown in Figure \ref{fig:stacks}, the average UVJ-quiescent galaxy from LEGA-C has \oii\ present in its spectrum.  This is true regardless of if the \oii\ is detected significantly in LEGA-C, i.e. if the EW is in excess of 1.5 \AA.  It is natural to ask if we would still detect \oii\ in the stack of individual low-EW objects using a different detection threshold.  A lower threshold would result in more contamination in the sample of ``detections" due to more spurious, low-S/N objects being classified as detections.  Conversely, a lower threshold would not result in additional contamination from true detections in the sample of ``non-detections'' and, moreover, any contaminants remaining in this sample would result in larger bootstrap uncertainties due to the smaller overall number of sources.

\begin{figure}
\begin{center}
\includegraphics[width=.45\textwidth]{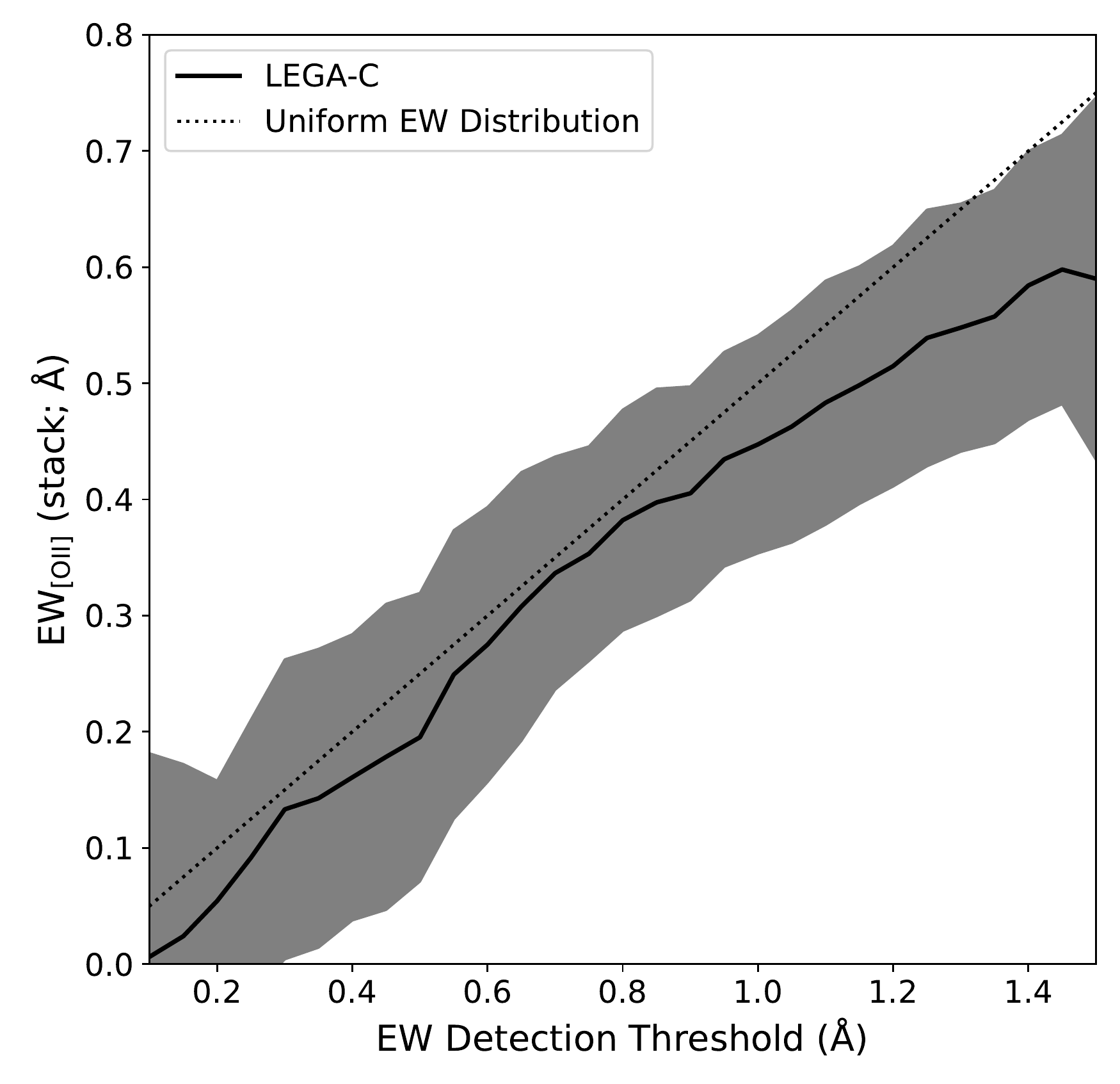} 
\caption{\oii\ EW determined from the median-stacked spectrum of ``non-detections" as a function of the EW threshold used to define an individual detection.  The shaded region denotes 3-$\sigma$ uncertainties, determined from bootstrap resampling of the resulting sample.  The similarity with the 2:1 line (dotted), which would be expected for a uniform distribution of EWs, implies that \oii\ is ubiquitous in the sample even below our EW completeness limit.}
\label{fig:ewthresh}
\end{center}
\end{figure}

To demonstrate the effect of the EW threshold on the measured \oii\ line in the stacked spectrum, we vary this threshold and repeat the stacking procedure outlined in Section \ref{sec:sample}.  Figure \ref{fig:ewthresh} shows the resulting \oii\ EW from the stack of individual low-EW sources as a function of the threshold.  We see that the median \oii\ EW from the stack decreases as the detection threshold is decreased.   However, we find that \oii\ is detected in the stack at greater than 3-$\sigma$ even when only including sources that individually have EW values less than 0.3 \AA.  The similarity with the trend compared to that expected from a uniform distribution of EWs (dotted line), where the median stacked EW is equal to half of the EW threshold, provides further evidence that low-EW ($\lesssim$ 1.5 \AA) \oii\ emission is ubiquitous in the population of LEGA-C quiescent galaxies.

\subsubsection{Choice of SPS model}
\label{sec:sps}
As described in Section \ref{sec:pf}, we fit the stellar continuum of the galaxies with the empirically-based MILES models.  We could also have used the theoretical templates of stellar populations created using the FSPS package \citep{2009ApJ...699..486C,2012ApJ...747...69C} from C. Conroy (private communication; see also A. van der Wel et al., submitted), which offer higher spectral resolution ($\sigma = $ 12 km s$^{-1}$ compared to 70 km s$^{-1}$) that is better-matched to our observations ($\sigma_{\mathrm{instr}} = $ 35 km s$^{-1}$), or the BC03 models \citep{2003MNRAS.344.1000B}.  We choose the MILES models as our fiducial model as they produce the smallest residuals in the continuum-subtracted stacks ($\chi^2 =$ 1.49 compared to 4.85 for the Conroy-FSPS models and 7.14 for BC03 models in the region around \oii: $\lambda$ = 3650$-$3720 and 3750$-$3850 \AA).

\begin{figure}
\begin{center}
\includegraphics[width=.45\textwidth]{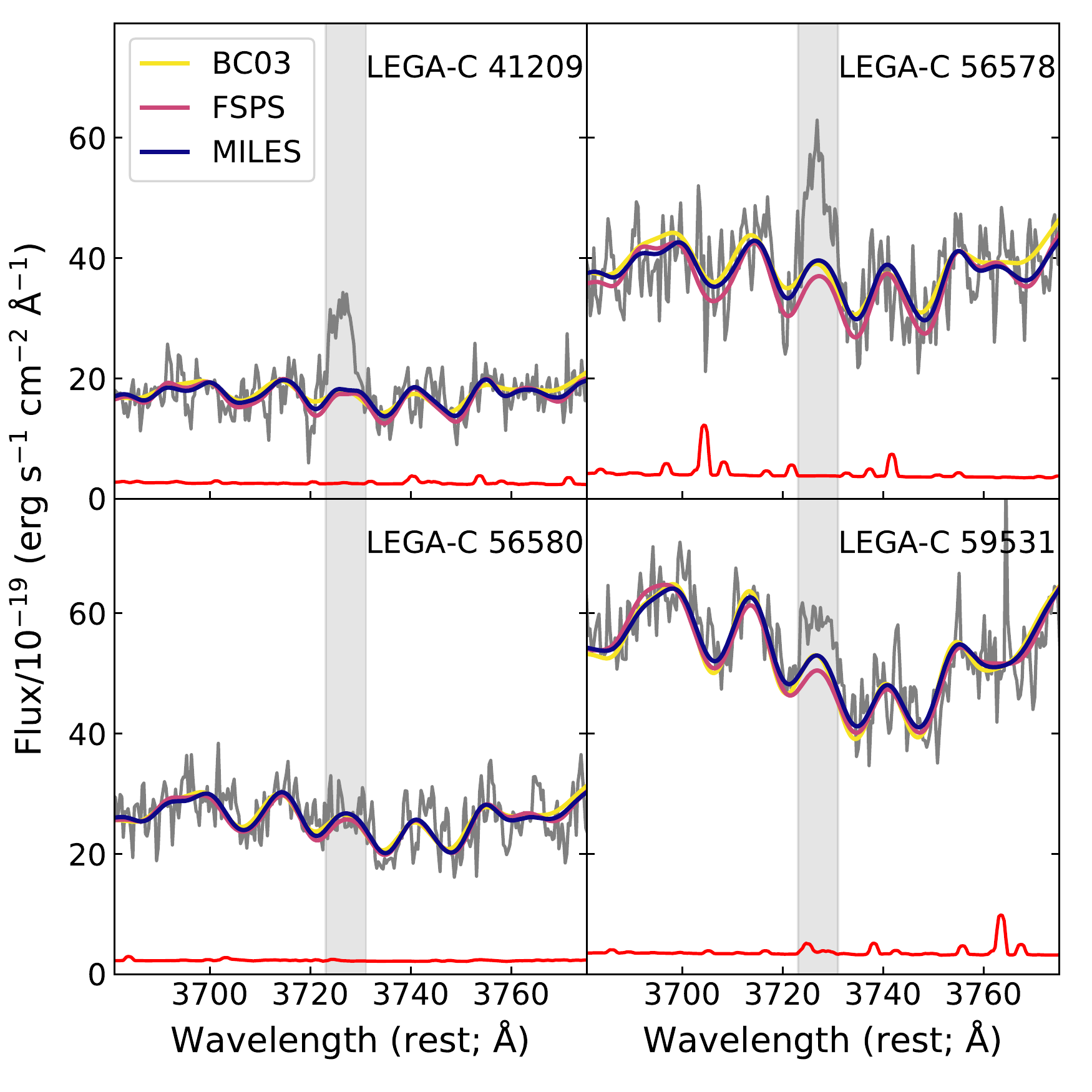}
\caption{Same as Figure \ref{fig:specs}, highlighting the difference between stellar population models that can be used to fit the stellar continua of LEGA-C galaxies: FSPS \citep{2009ApJ...699..486C,2012ApJ...747...69C}, BC03 \citep{2003MNRAS.344.1000B}, or MILES \cite[our default model;][]{2010MNRAS.404.1639V}.  Overall, correcting for absorption around the position of \oii\ is critical to obtaining accurate fluxes and EWs, but the specific choice of model is of secondary importance.}
\label{fig:sps}
\end{center}
\end{figure}

While these (and other) models offer differing predictions for e.g. the continuum shape and strength of absorption features relating to stellar evolution phases that may be important producers of the ionizing photons we are observing \cite[e.g. AGB stars;][]{2006ApJ...652...85M}, the main features we are concerned with here are the high-order Balmer absorption features around the position of \oii.  Figure \ref{fig:sps} shows these three models fit to the same galaxies as in Figure \ref{fig:specs}, again using \texttt{pPXF}.  In all cases, the uncertainty in the individual emission line flux measurement is larger than the systematic uncertainty in the continuum fit stemming from the choice of stellar population model.  As such, we conclude that the choice of stellar population model does not have a significant effect on the overall conclusions of this work, namely that most UVJ-quiescent galaxies in the LEGA-C survey have \oii\ emission regardless of the model we use to correct for the underlying Balmer absorption.

\section{Comparison to the Local Universe}
\label{sec:gama}

We derive a sample of galaxies in the local Universe from the GAMA survey, data release 3 \citep{2011MNRAS.413..971D,2015MNRAS.452.2087L,2018MNRAS.474.3875B}.  At redshifts $z <$ 0.09, the main GAMA survey is spectroscopically more than 99\% complete above a stellar mass of 10$^{10}$ \msol, matching the LEGA-C mass range \citep{2011MNRAS.418.1587T}.  The AAOmega spectrograph used in GAMA offers spectral coverage of \oii\ for all galaxies \citep{2006SPIE.6269E..0GS}.

We differentiate quiescent and star-forming galaxies using the low-$z$ UVJ criteria of \citet{2011ApJ...735...86W}, applying \texttt{EAZY} to the GAMA ``LAMBDAR'' photometry \citep{2016MNRAS.460..765W} and using the catalog spectroscopic redshifts.  Importantly, this photometry covers the restframe $J$-band directly (cf. the $ugriz$ photometry of SDSS).  Emission lines are measured using the same methodology as presented in Section \ref{sec:pf} and the 1D spectra are absolute flux-calibrated to the observed $r$-band.  
In total, we consider 340 GAMA sources with redshifts below 0.09 and stellar masses above 10$^{10.2}$ \msol\ (to better match LEGA-C).  The median redshift for the objects in this sample is $\left<z\right>$ = 0.077.  While GAMA has spectral coverage of \oii\ for all galaxies, the efficiency is quite low at wavelengths shorter than 4000 $\mathrm{\AA}$ (observed).  Therefore, the sensitivity to low EW values is lower than LEGA-C: at an EW threshold of 1.5 $\mathrm{\AA}$, only 75 (66) percent of GAMA (quiescent) galaxies have \oii\ measurements with a significance $>$ 3-$\sigma$.  Only at EWs above 5.1 $\mathrm{\AA}$ are 90 percent of the GAMA  \oii\ measurements $>$ 3-$\sigma$.

\begin{figure*}
\begin{center}
\includegraphics[width=.9\textwidth]{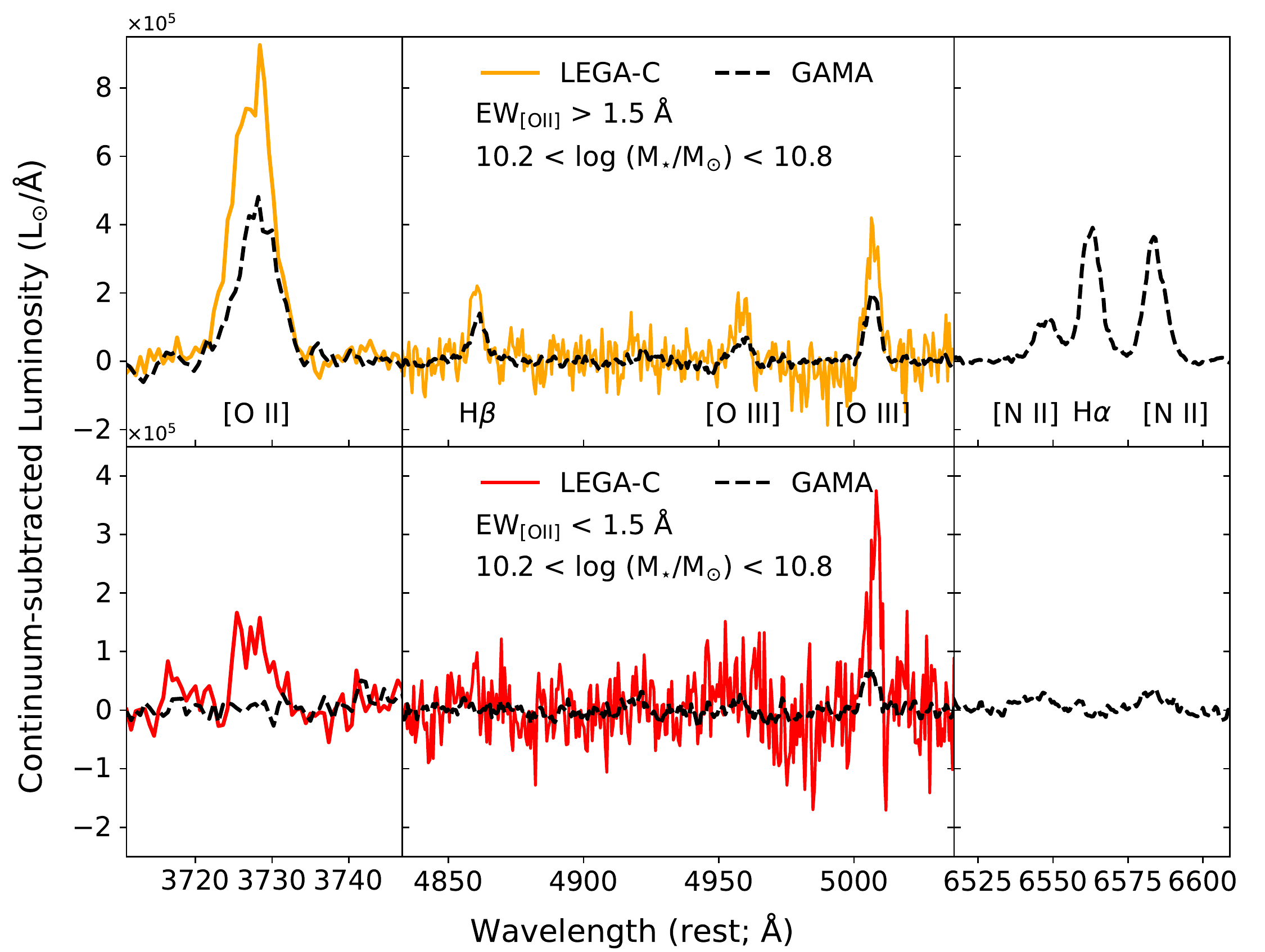}
\caption{Median, continuum-subtracted stacked spectra for LEGA-C (solid) and GAMA (dashed) quiescent galaxies, split by their \oii\ EWs into high-EW (top; $n =$ 97 LEGA-C/66 GAMA) and low-EW (bottom; $n =$ 70/63).  While the GAMA stack has a larger \oii\ EW for the EW $>$ 1.5 $\mathrm{\AA}$ galaxies (5.9 versus 4.2 \AA), this is driven primarily by an evolution in the continuum luminosities: LEGA-C galaxies have an \oii\ luminosity that is approximately a factor of 3 larger than GAMA galaxies (upper-left panel).  For galaxies with individual measurements less than our completeness limit of 1.5 \AA, only the LEGA-C stack still shows \oii\ emission on average (bottom-left panel).  The 5-$\sigma$ upper limit for the EW of the GAMA stack of low-EW sources is 0.1 $\mathrm{\AA}$. This implies that, although the total fraction of quiescent galaxies with \oii\ EWs above 1.5 $\mathrm{\AA}$ and its corresponding trend with mass is similar for both surveys, low-EW \oii\ emission is only ubiquitous in LEGA-C galaxies.  This is in contrast with \oiii, which is detected in both stacks: 0.2 $\pm$ 0.04 \AA\ from LEGA-C and 0.3 $\pm$ 0.04 \AA\ from GAMA.  GAMA provides spectral coverage of the \ha\ and \nii\ emission features as well (right panels), with ratios indicative of contributions from sources other than star formation (see Section \ref{sec:discussion}).}
\label{fig:gamastack}
\end{center}
\end{figure*}

In Figure \ref{fig:gamastack} we show the median-stacked, continuum-subtracted LEGA-C and GAMA spectra for quiescent galaxies with (orange) and without (red) individual \oii\ EWs in excess of 1.5 $\mathrm{\AA}$.  Even when stacking 63 low-EW sources from GAMA we do not see \oii\ and can place a 5-$\sigma$ upper limit to the EW of 0.1 $\mathrm{\AA}$.  We do, however, observe \oiii\ in the GAMA stack containing galaxies with individual detections of \oii.  This is different from what we see when looking at the LEGA-C galaxies, where \oii\ is detected in both stacks (in addition to \oiii; Figures \ref{fig:specs} and \ref{fig:ewthresh}).

\begin{figure*}
\begin{center}
\includegraphics[width=.9\textwidth]{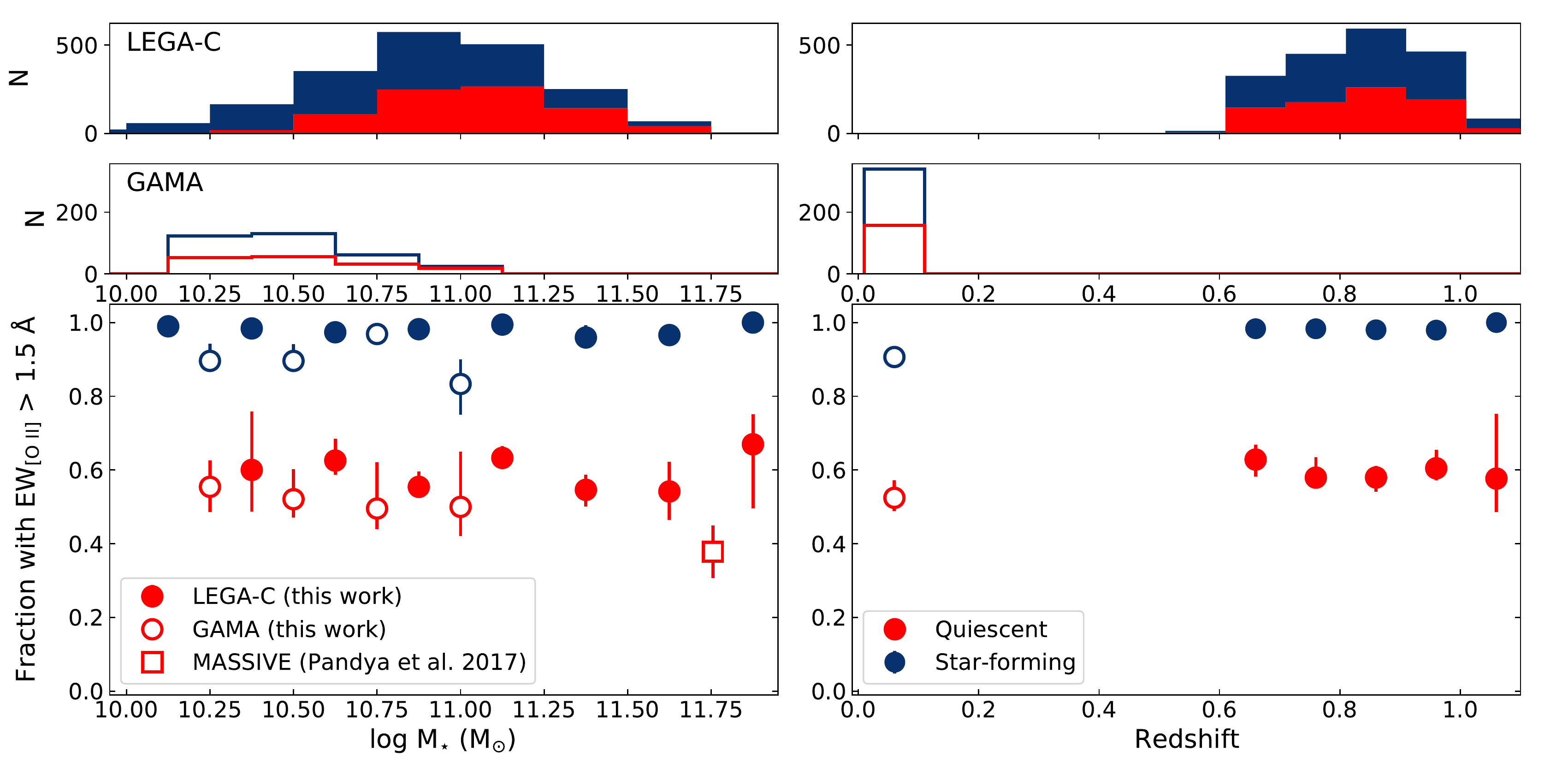}
\caption{Fraction of galaxies with \oii\ EW in excess of 1.5 \AA\ in bins of stellar mass (left) and redshift (right).  Error bars are based on bootstrapping the samples, except in the case of MASSIVE (at $z\approx$ 0) where they are Poisson.  The incidence rate of \oii\ emission does not strongly depend on stellar mass, although GAMA has a very limited number of quiescent galaxies above 10$^{11}$ \msol.  However, the 1.7-$\sigma$ tension between LEGA-C and MASSIVE at stellar masses above 10$^{11.5}$ \msol\ and the presence of \oii\ in the stacked spectra of low-EW LEGA-C sources implies that the increase in detection fraction with redshift could be a physical effect.}
\label{fig:massfrac}
\end{center}
\end{figure*}

Figure \ref{fig:massfrac} shows how the incidence rate of \oii\ emission varies in bins of stellar mass for the star-forming (blue) and quiescent (red) populations in  LEGA-C and GAMA.  Stellar masses (and star formation rates) for LEGA-C come from \texttt{Prospector}, as described in Section \ref{sec:data}.  There is no clear decrease in the \oii\ detection fraction in quiescent galaxies as a function of stellar mass in LEGA-C, in contrast to the result from \citet{2013AA...558A..61M}, who find that the fraction of UVJ-quiescent galaxies that display \oii\ emission ($>$ 5 \AA) decreases with stellar mass.  This effect is most pronounced below 10$^{10.25}$ \msol, which is not sampled in LEGA-C.  We do, however, observe a higher median \oii\ EW in galaxies with masses between 10$^{10.25}$ and 10$^{10.75}$ \msol\ compared to galaxies with masses above 10$^{10.75}$ \msol\ (2.4 versus 2.3 \AA), in agreement with \citet{2013AA...558A..61M}.

Further, it is difficult to assess any difference for stellar masses above 10$^{11}$ \msol\ due to the relative lack of galaxies in the GAMA sample.  In order to supplement our results from GAMA, we turn to results from the MASSIVE survey from \citet{2017ApJ...837...40P}, a $z\approx$ 0 survey targeting the most massive quiescent galaxies in the local Universe.  Compared to MASSIVE, we observe a detection fraction that is tentatively higher in LEGA-C at the 1.7-$\sigma$ level when restricting to stellar masses above 10$^{11.5}$ \msol. 

In order to assess any potential redshift evolution, we focus on the stellar mass range of maximal overlap between GAMA and LEGA-C, 10$^{10.2}-$10$^{10.8}$ \msol.  In this range, 58$\pm$6 percent of LEGA-C quiescent galaxies have \oii\ in excess of 1.5 \AA, compared to 51$\pm$6 percent of GAMA galaxies \cite[see also][]{2006ApJ...648..281Y}.  We see a non-unity detection fraction for star-forming galaxies in GAMA, even though they should theoretically all have prominent \oii\ emission \citep{2001ApJ...551..825J,2004AJ....127.2002K}.  Such a systematic offset could be explained by the lower S/N of the GAMA spectra at the wavelength of \oii, where real \oii\ emission could fall below our EW threshold due to noise.  However, the non-detection of \oii\ in the GAMA stack of low-EW quiescent galaxies (bottom panel of Figure \ref{fig:gamastack}) implies that we could still be observing a physical difference between the samples even though the detection fractions are formally consistent.  This could be a real evolution in the detection fraction with redshift (see the right panel of Figure \ref{fig:massfrac}), related to an ionizing source that is more prevalent at earlier cosmic times or in younger galaxies (cf. the median D$_n$4000 from the GAMA galaxies of 1.78 versus 1.66 for LEGA-C), or to an evolution in the characteristic EW of \oii\ for a fixed ionizing source.  In this latter case, the evolution could be due to a different ionizing spectrum, a different star-gas geometry, or a different amount of gas in the galaxy \cite[e.g.][]{2018ApJ...860..103S}.

\subsection{Discussion}
\label{sec:discussion}

\begin{figure}
\begin{center}
\includegraphics[width=.45\textwidth]{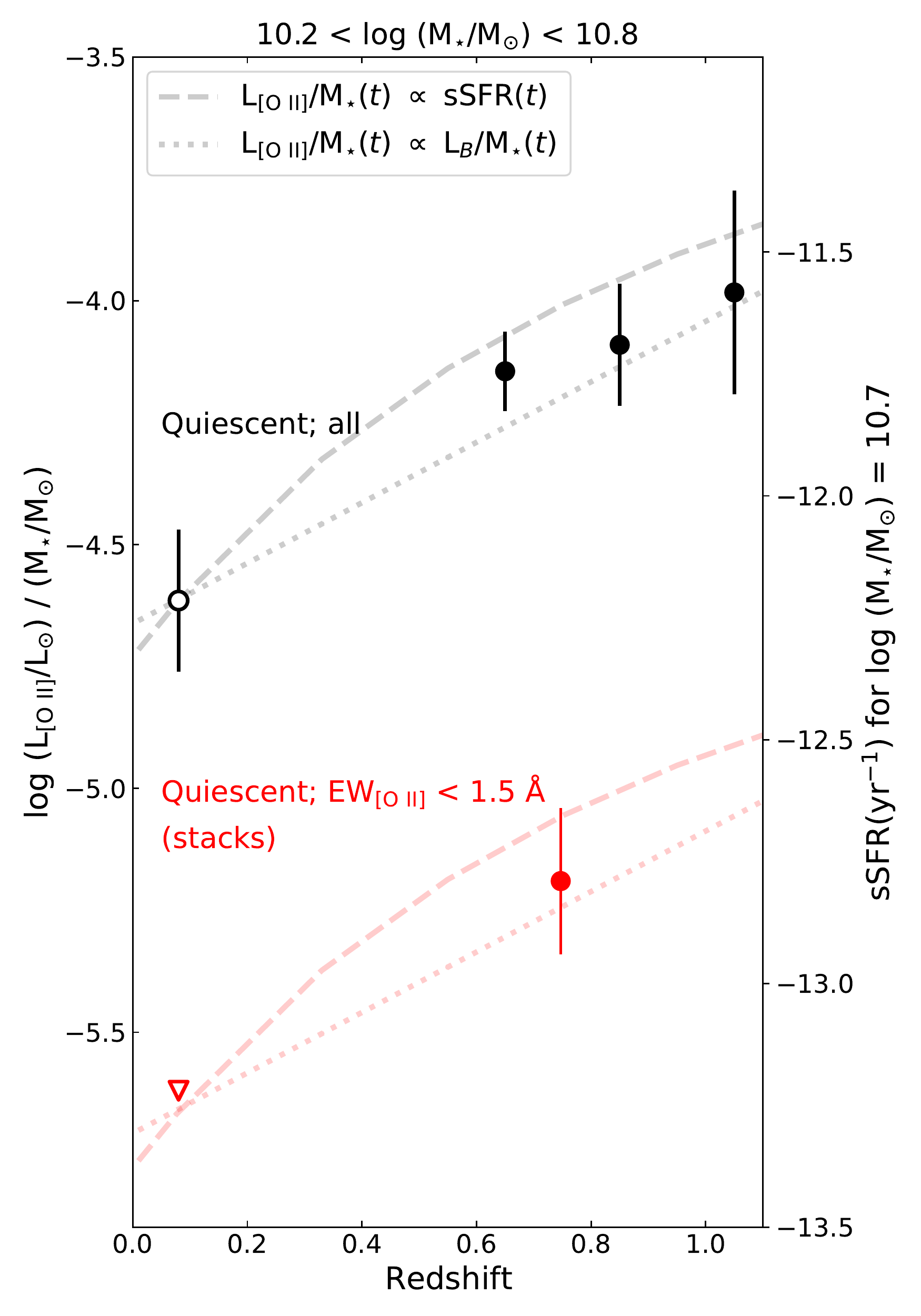}
\caption{Redshift evolution of the median ``specific \oii\ luminosity," the measured \oii\ luminosity normalized to the stellar mass of the galaxy, for LEGA-C (filled) and GAMA (open) galaxies with stellar masses in the range 10$^{10.2}-$10$^{10.8}$ \msol.  Results from stacked spectra for individual quiescent galaxies with \oii\ EWs below 1.5 \AA\ are shown in red (Figure \ref{fig:gamastack}; the GAMA upper-limit is 5-$\sigma$).  Errors on the medians are derived by random resampling of the galaxies in each redshift bin.  For all quiescent galaxies, the specific \oii\ luminosity increases strongly with redshift.  The right $y$-axis shows a representative sSFR value assuming M$_{\star} =$ 10$^{10.7}$ \msol\ and all \oii\ comes from star formation, assuming the \citet{2004AJ....127.2002K} calibration.  The redshift evolution for the full sample of quiescent galaxies matches the evolution of specific star formation rates in the Universe \cite[][dashed line]{2014ApJS..214...15S}, as noted in \citep{2014ApJ...796...35F}, but it also matches the evolution in $B$-band stellar mass-to-light ratios \cite[][dotted line]{2005ApJ...631..145V}.  For the $EW < 1.5$ \AA\ sample, the evolution could be much faster although the nature of the upper-limit from GAMA makes a firm conclusion difficult to make.}
\label{fig:loii}
\end{center}
\end{figure}

Diagnosing the source of the disagreement in the stacks with EWs alone is difficult because an elevated EW could be due to a larger line flux and/or a fainter stellar continuum level, which are both related to the age of the galaxy.  As the general population of quiescent galaxies ages significantly from $z\approx0.85$ (LEGA-C) to $z\approx0.1$ (GAMA), galaxies  fade at fixed stellar mass \cite[e.g.][]{2005ApJ...633..174T}.  This corresponding drop in the stellar continuum level would mean that, at a fixed \oii\ luminosity, we would expect to observe an increased \oii\ EW in GAMA compared to LEGA-C in the same mass range: based on the evolution of the $B$-band mass-to-light ratios, this would be a factor of 3.0 \citep{2005ApJ...631..145V}.  We do see an observed evolution in the \oii\ EWs from LEGA-C and GAMA for the EW $>$ 1.5 \AA\ sub-samples (4.2$\pm$0.42 and 5.9$\pm$0.60 \AA, respectively), but the evolution is slower than would be expected from an evolving continuum mass-to-light ratio without any evolution in the \oii\ luminosities.  Furthermore, we see a \textit{higher} EW in LEGA-C for the stacked spectra of EW $<$ 1.5 \AA\ galaxies and a higher median \oii\ EW for the overall population of quiescent galaxies (2.4 vs 1.5 \AA).  Therefore, LEGA-C quiescent galaxies must have higher \oii\ luminosities per unit stellar mass than GAMA galaxies.  This is visible in Figure \ref{fig:loii}, where we compare the observed ``specific \oii\ luminosity," i.e. the \oii\ luminosity normalized to the stellar mass of the galaxy.  The typical value for a LEGA-C quiescent galaxy is 0.5 dex larger than that for a GAMA galaxy (black points), 10$^{-4.1}$ L$_{\odot}$ M$_{\odot}^{-1}$ compared to 10$^{-4.6}$ L$_{\odot}$ M$_{\odot}^{-1}$, matching the rate of evolution in the stellar mass-to-light ratios.  The distributions are different to $>$ 10-$\sigma$, with qualitatively similar behavior when we look at the quiescent galaxies with low-EW or no \oii\ (red points). 

As shown in e.g. \citet{2014ApJ...796...35F} and \citet{2019ApJ...877..140L}, however, the specific star formation rates of quiescent galaxies are higher at higher redshift, albeit always much lower than what is observed in star-forming galaxies.  Since \oii\ is a (metallicity- and excitation-dependent) star formation rate indicator, an elevated \oii\ luminosity in LEGA-C could be indicative of more recent or ongoing star formation in the quiescent population: if all of the \oii\ luminosity comes from star formation, the median star formation rates for the \oii-detected and non-detected LEGA-C sources are 1.0 and 0.09 M$_{\odot}$ yr$^{-1}$, respectively \citep{2004AJ....127.2002K}.  If this were the case, we could expect to see an evolution in the specific \oii\ luminosity with redshift that matches what is seen for star-forming galaxies: Figure \ref{fig:loii} shows that the redshift evolution in the specific \oii\ luminosity for all quiescent galaxies also matches the sSFR evolution and the evolution in the continuum stellar mass-to-light ratios.  For the lowest EW quiescent galaxies, the strong redshift evolution (i.e. \oii\ is not detected in the GAMA stack but is detected in the LEGA-C stack) is also consistent with the same scaling although the fact that GAMA only provides an upper-limit hampers a more detailed comparison.

We can combine the redshift evolution in the specific \oii\ luminosity with spectral information from GAMA to provide some insight into the potential production mechanism(s) for \oii.  Specifically, the spectral coverage of GAMA (3700$-$8800 \AA) allows us to use line ratios to determine the dominant source of the ionizing photons for those samples; we do not have spectral coverage of \ha\ or \nii\ for the LEGA-C galaxies (right panels of Figure \ref{fig:gamastack}).  For the EW $<$ 1.5 \AA\ sample from GAMA, the log \oiii/\hb\ lower limit is 0.50 and the log \nii/\ha\ lower limit is 0.37 (3-$\sigma$, although this is quite uncertain as \ha\ is undetected and \nii\ only has a S/N of 2.2).   These line ratios, although only providing weak limits, imply that the stacked spectrum can likely be classified a ``LIER" or ``LINER," as is commonly observed in red, early-type systems at low- and intermediate-$z$ \cite[e.g.][]{2006ApJ...648..281Y,2013MNRAS.429.2212M,2010ApJ...716..970L,2012ApJ...747...61Y,2013AA...558A..43S,2016MNRAS.461.3111B}.  In the stack based on the EW $>$ 1.5 \AA\ sample from GAMA, the log \oiii/\hb\ ratio of 0.14 $\pm$ 0.13 and log \nii/\ha\ ratio of -0.066 $\pm$ 0.17 mean that spectrum can be classified as a ``composite" or a combination of star formation and either LIER or an active galactic nucleus \citep{2006MNRAS.372..961K}.  We note that even though the 1$''$ VIMOS slits for LEGA-C typically probe a slightly larger physical region than the 2$\farcs$1 GAMA fibers ($\approx$ 2.3 compared to 1.3 $R_e$), results from spatially-resolved spectroscopy suggest that both line EW gradients and line ratio gradients are likely to be quite flat in these systems integrated out to these radii \citep{2016MNRAS.461.3111B}.  

The ``composite" nature of the emission line ratios from the stack of sources with EWs in excess of 1.5 \AA\ is supported by the difference between the \oii-predicted star formation rate and that predicted by the \hb\ luminosity: a SFR of 1 \msol\ yr$^{-1}$, consistent with the stacked \oii\ luminosity, would have resulted in an \hb\ flux of 190$\times$10$^{-19}$ \cgs\ at solar metallicity without dust \citep{2011ApJ...737...67M}, or a factor of 7.9 larger than that measured in Table \ref{tab:lines}.  We can therefore rule-out star formation as the \textit{sole} source of ionizing photons in the high-EW galaxies \cite[modulo the effects of metallicity, age, or dust attenuation;][]{2004AJ....127.2002K}, although a fractional contribution of star formation to the ionizing photon budget could be driving the redshift evolution.  Contrarily, for the low-EW subsample the predicted \hb\ flux from the \oii-based SFR of 0.09 \msol\ yr$^{-1}$ is 17$\times$10$^{-19}$ \cgs, consistent with the stacked upper limit (Table \ref{tab:lines}).  Together with the implied redshift evolution in the specific \oii\ luminosity, we likely cannot conclusively rule-out star formation in these galaxies without better constraints from high-resolution spectra covering \ha\ and \nii\ for LEGA-C galaxies.

If the \oii\ in these galaxies were caused at least in part by star formation, we note that it would have to occur on relatively short timescales so as not to strongly affect the H$\delta$ strength (which is only sensitive to star formation after 50$-$100 Myr; see also Figure \ref{fig:d4000}) or the broadband magnitudes and colors \cite[see][and submitted]{2018ApJ...855...85W}.  Since the total UV and IR luminosities of these galaxies are still likely to be dominated by hot, evolved stars and not stars formed very recently \citep{2007ApJ...671..285T,2014ApJ...796...35F,2019ApJ...877..140L}, care must be taken to derive ``corrected" star formation rates that are more indicative of the near-instantaneous formation of massive stars: \oii\ (or \hb) luminosities alone are not sufficient.  Detailed modeling, including contributions from star formation as well as Seyfert/AGN and evolved stellar populations are required in order to disentangle the various contributions to the observed samples \citep{2018MNRAS.481..476Y}.  We defer such an analysis to future work.

\section{Summary and Conclusions}
\label{sec:conclusions}

Using extremely deep (S/N $\approx$ 20 \AA$^{-1}$) spectroscopy from the mass-complete LEGA-C survey, we assess the prevalence of \oii\ $\lambda\lambda$3727,3729 emission in $z \approx 0.85$ galaxies that are classified as quiescent (i.e. not star-forming) according to the UVJ rest-frame color-color diagram.  Our primary findings are as follows:
\begin{itemize}
\item LEGA-C galaxies with stellar masses in excess of $10^{10.2}$ \msol, selected to be quiescent according to the UVJ-criteria, ubiquitously show \oii\ emission in their spectra.  We see this in both individual and stacked spectra.  \oii\ is the strongest emission line observed in the stacked spectra, although other emission lines such as \ha\ (and \nii), \oiii, and \hb\ are also present (Section \ref{sec:sample} and Figure \ref{fig:stacks}).
\item We observe correlations between the strength/detection fraction of \oii\ and both UVJ colors and \hda/D$_n$4000 spectral indices, although significant overlap in the distributions prevent us from making an accurate \textit{a priori} prediction for the presence/absence of \oii\ in a given galaxy (Figures \ref{fig:uvj}, \ref{fig:sed}, and \ref{fig:d4000}; Section \ref{sec:diffs}).
\item The ubiquitous \oii\ emission at $z\approx0.85$ contrasts with our results at $z\approx0.1$ from GAMA, where stacks of the individual low-EW galaxies do not show signatures of \oii\ (Section \ref{sec:gama} and Figure \ref{fig:gamastack}).
\item We see no strong evidence for an evolving incidence rate of EW $>$ 1.5 \AA\ \oii\ emission with stellar mass at all redshifts probed here (Figure \ref{fig:massfrac}).  We do, however, observe significant redshift evolution in the \oii\ luminosity (Figure \ref{fig:gamastack}), also when normalized by the stellar mass of the galaxy (Figure \ref{fig:loii}).
\item We can classify the GAMA galaxies 
with \ewoii\ $<$ ($>$) 1.5 \AA\ as LIERs/LINERs (``composites") according to their line ratios.  In both cases, the ubiquitous \oii\ emission and the strength of the redshift evolution implies that star formation could potentially contribute fractionally to the observed \oii\ luminosity (Section \ref{sec:discussion}).
\end{itemize}

Further work is required to understand the source(s) of the ionizing photons seen in these quiescent galaxies.  If the \oii\ emission comes predominantly from ongoing star formation, we would expect to see correlations between the strength of \oii\ and the strength of Balmer emission lines like \ha\ and \hb, albeit with the caveat that the \oii-SFR relationship depends on the metallicity and the dust attenuation \cite[e.g.][]{2001ApJ...551..825J}.  Detailed SED modeling can take old stellar populations into account when fitting the observed UV and IR luminosities \citep{2019ApJ...877..140L}, which will allow us to compare the observed \oii\ luminosities with the current SFR even when individual Balmer lines are not detected.  The strength of \oii\ can be combined with age-sensitive spectral features such as H$\delta$ absorption and the D$_n$4000 index in order to compare with predictions for the evolution in the strength of \oii\ with time from e.g. stellar population models that include contributions from evolved stars \citep{Byler_2019}.

Spatially-resolved 2D spectroscopy from LEGA-C as well as follow-up spectroscopy covering \ha\ and \nii\ will also allow us to investigate the incidence rate of AGN in the quiescent population and the contribution of maintenance-mode feedback to the ionizing photon budget \citep{2013ApJ...770L..39W}.  Indeed, both \citet{2017AA...599A..95G} and \citet{2017ApJ...838...94W} utilize spatial information to understand the role of AGN in the production of \oii\ photons at $z = 1.6$ and $1.2$, respectively.  \citet{2017ApJ...838...94W} combine morphological cuts with \oii\ (non-)detections and X-ray data to conclude that AGN are unlikely to contribute a significant amount to the integrated \oii\ luminosities, also finding a lower \oii\ detection fraction in spatially-compact sources (albeit with shallower spectra than those used in this study).  \citet{2017AA...599A..95G} find higher EWs in the wings of the stacked radial \oii\ emission line profile compared to the core (11 versus 5 \AA) even with similar stellar continua.  They interpret this as evidence for young stars being the primary contributor of \oii\ photons in their ($BzK$-selected) quiescent galaxies.  We note that both of these surveys find typical \oii\ EWs in quiescent galaxies of 5 \AA\ (i.e. larger than we find in LEGA-C), which could be indicative of higher SFRs at these higher redshifts \cite[e.g. 4.5 \msol\ yr$^{-1}$ in \citeauthor{2017AA...599A..95G} \citeyear{2017AA...599A..95G}; see also][]{2014ApJ...796...35F}.  The quality of the LEGA-C spectra is high enough to perform spatially-resolved studies on individual objects (Straatman et al. in prep).  This work will provide a more detailed look into the production mechanisms for the ubiquitous \oii\ emission in $z\approx0.85$ quiescent galaxies.

\begin{acknowledgements}

The authors would like to thank Nell Byler, Jarle Brinchmann, and the anonymous referee for insightful comments and discussions.  This project has received funding from the European Research Council (ERC) under the European Union's Horizon 2020 research and innovation program (grant agreement No. 683184).

Based on observations made with ESO Telescopes at the La Silla Paranal Observatory under program IDs 179.A-2004 (the VISTA Kilo-degree Infrared Galaxy Survey), 179.A-2005 (the Deep/Ultra-Deep Near-IR Survey of the COSMOS Field), and 194-A.2005 (the LEGA-C Public Spectroscopy Survey). GAMA is a joint European-Australasian project based around a spectroscopic campaign using the Anglo-Australian Telescope. The GAMA input catalogue is based on data taken from the Sloan Digital Sky Survey and the UKIRT Infrared Deep Sky Survey. Complementary imaging of the GAMA regions is being obtained by a number of independent survey programs including GALEX MIS, VST KiDS, VISTA VIKING, WISE, Herschel-ATLAS, GMRT and ASKAP providing UV to radio coverage. GAMA is funded by the STFC (UK), the ARC (Australia), the AAO, and the participating institutions. The GAMA website is http://www.gama-survey.org/.

\end{acknowledgements}

\facilities{AAT, ESO:VISTA, HST, VLT:Melipal}
\software{Astropy \citep{astropy:2013, astropy:2018}, EAZY \citep{2008ApJ...686.1503B}, Matplotlib \citep{Hunter:2007}, MPDAF \citep{2016ascl.soft11003B}, NumPy \citep{2020NumPy-Array}, pPXF \citep{2012ascl.soft10002C}, Prospector \citep{2019ascl.soft05025J}, pyFSPS \citep{ben_johnson_2021_4577191}, SciPy \citep{2020SciPy-NMeth}, TOPCAT \citep{2005ASPC..347...29T}}

\bibliographystyle{aa} 

\bibliography{legac.bib}

\end{document}